\documentclass[journal]{IEEEtran}
\usepackage{amsmath,graphicx, hyperref,multirow,epsfig}
\usepackage{pgfplots}
\usepackage{pifont}

\usepackage{tikz}

\ifCLASSINFOpdf
\else
\fi
\usepackage{algorithmic}
\begin{document}
%
\title{Lossless Image and Intra-frame Compression with Integer-to-Integer DST}
%
%
%

\author{Fatih Kamisli,~\IEEEmembership{Member,~IEEE}
\thanks{F. Kamisli is with the Department of Electrical and Electronics Engineering at the Middle East Technical University, Ankara, Turkey.}

}

\maketitle

\begin{abstract}
Video coding standards are primarily designed for efficient lossy compression, but it is also desirable to support efficient lossless compression within video coding standards using small modifications to the lossy coding architecture. A simple approach is to skip transform and quantization, and simply entropy code the prediction residual. However, this approach is inefficient at compression. A more efficient and popular approach is to skip transform and quantization but also process the residual block with DPCM, along the horizontal or vertical direction, prior to entropy coding. This paper explores an alternative approach based on processing the residual block with integer-to-integer (i2i) transforms. I2i transforms can map integer pixels to integer transform coefficients without increasing the dynamic range and can be used for lossless compression. We focus on lossless intra coding and develop novel i2i approximations of the odd type-3 DST (ODST-3). Experimental results with the HEVC reference software show that the developed i2i approximations of the ODST-3 improve lossless intra-frame compression efficiency with respect to HEVC version 2, which uses the popular DPCM method, by an average 2.7\% without a significant effect on computational complexity. 
\end{abstract}



%
\IEEEpeerreviewmaketitle

\section{Introduction}
\label{sec:intro}


Video coding standards are primarily designed for efficient lossy compression, but it is also desirable to support efficient lossless compression within video coding standards. However, to avoid increase in the system complexity, lossless compression is typically supported using small modifications to the lossy coding architecture.
 
Lossy compression in modern video coding standards, such as HEVC \cite{HEVC} or H.264 \cite{Luthra264}, is achieved with a block-based approach. First, a block of pixels are predicted using pixels either from a previously coded frame (inter prediction) or from previously coded regions of the current frame (intra prediction). The prediction is in many cases not sufficiently accurate and in the next step, the block of prediction error pixels (residual) are computed and then transformed to reduce remaining spatial redundancy. Finally, the transform coefficients are quantized and entropy coded together with other relevant side information such as prediction modes. 

To support also lossless compression within the block-based lossy coding architecture summarized above, the simplest approach is to just skip the transform and quantization steps, and directly entropy code the prediction residual block. This approach is indeed used in HEVC version 1 \cite{HEVC}. While this is a simple and low-complexity approach, it is well known that  prediction residuals are not sufficiently decorrelated in many regions of video sequences and directly entropy coding a prediction residual block is inefficient at compression. Hence, a large number of approaches have been proposed to develop more efficient lossless compression methods for video coding.

A more efficient and popular approach is to skip transform and quantization but process the residual block with differential pulse code modulation (DPCM) prior to entropy coding \cite{sulivanDPCM,cross}. While there are many variations of this approach \cite{sulivanDPCM,cross,IRDPCM,ARDPCM}, video coding standards HEVC and H.264 include the simple horizontal and vertical DPCM due to their low complexity and reasonable compression performance. 


This paper explores an alternative approach for lossless compression within video coding standards. Instead of DPCM, integer-to-integer (i2i) transforms are used to process the residual block. I2i transforms can map inputs that are on a uniform discrete lattice to outputs on the same lattice and are invertible \cite{goyal_i2i}. In other words, i2i transforms can map integer pixels to integer transform coefficients. Note however that unlike the integer transforms used in HEVC for lossy coding \cite{HEVCtr}, i2i transforms do not increase the dynamic range at the output and can therefore be easily employed in lossless coding. While there are many papers that employ i2i approximations of the discrete cosine transform (DCT) in lossless image compression \cite{binDCT}, we could not come across a work which explores i2i transforms for lossless compression of prediction residuals in video coding, or particularly in H.264 or HEVC. 

This paper focuses on lossless compression for intra coding. For lossless inter coding, some of our preliminary results are provided in \cite{kamisli2016losslessTR}. In lossy intra coding, it is known that a hybrid separable 2D transform based on the odd type-3 discrete sine transform (ODST-3) and the DCT \cite{Chuo_dst,Han_dst} or simply a 2D ODST-3 \cite{HEVC} provides improved compression performance over the traditional 2D DCT at transform coding block-based spatial prediction residuals. 
While the literature includes great previous research on i2i DCTs \cite{binDCT,i2iSERM,i2iLU}, 
we could not find any i2i approximations of the ODST-3. Therefore in this paper, we first explore the design of i2i approximations of the ODST-3 and then provide lossless intra-frame compression results with the developed i2i approximations of the ODST-3. Our experimental results performed using the HEVC reference software indicate that using the developed i2i approximations of ODST-3, the lossless intra-frame compression of HEVC version 2, which uses the popular DPCM method along the horizontal or vertical direction, can be improved by an average $2.7\%$ without significant complexity increase.



The remainder of the paper is organized as follows. In Section \ref{sec:pr}, a brief overview of related previous research on lossless video compression is provided. Section \ref{sec:i2i} discusses i2i transforms and their design based on plane rotations and the lifting scheme. Section \ref{sec:i2idst} presents a framework for designing computationally efficient i2i approximations of the ODST-3. Section \ref{sec:expres} presents experimental results with the designed i2i approximations of the ODST-3 within HEVC and compares them with those of HEVC version 1 and 2. Finally, Section \ref{sec:conc} concludes the paper. Note that some preliminary results of this work were presented in \cite{kamisli2016losslessTR,kamisli2016losslessDST}.


\section{Previous Research on Lossless Video Compression}
\label{sec:pr}

One of the simplest methods to support lossless compression within video codecs primarily designed for lossy coding is to just skip the transform and quantization steps, and directly entropy code the prediction residual block. This approach is indeed used in HEVC version 1 \cite{HEVC}. While this is a low-complexity approach, it is inefficient at compression since prediction residuals are typically not well decorrelated. Hence, a large number of approaches have been proposed to develop more efficient lossless compression methods for video coding. These approaches can be categorized into three groups, which we briefly review as follows.

\subsection{Methods based on residual DPCM}
\label{ssec:rdpcm}
The first group of methods are based on processing the residual blocks, obtained from the block-based spatial or temporal prediction of the lossy coding architecture, with differential pulse code modulation (DPCM) prior to entropy coding and are typically called residual DPCM (RDPCM) methods. There are many variations of RDPCM methods in the literature for both lossless intra and inter coding \cite{sulivanDPCM,cross,IRDPCM,ARDPCM}. RDPCM methods process the prediction residual block with some specific pixel-by-pixel prediction method, which is typically the distinguishing feature among the many RDPCM methods. 

One of the earliest RDPCM methods was proposed in \cite{sulivanDPCM} for lossless intra coding in H.264. Here, after the block-based spatial prediction is performed, a simple pixel-by-pixel differencing operation is applied on the residual pixels in only horizontal and vertical intra prediction modes. In the horizontal intra mode, from each residual pixel, its left neighbor is subtracted and the result is the RDPCM pixel of the block. Similar differencing is performed along the vertical direction in the vertical intra mode. Note that the residuals of other angular intra modes are not processed in \cite{sulivanDPCM} because directional pixel-by-pixel prediction with different interpolation for each angular prediction mode is required to account for the directional correlation of the residuals and the additional improvement in compression does not justify the complexity increase.

The same RDPCM method as in \cite{sulivanDPCM} is now included in HEVC version 2 \cite{HEVCv2,L0117} for intra and inter coding. In inter coding, RDPCM is applied either along the horizontal or vertical direction or not at all, and a flag is coded in each transform unit (TU) to indicate if it is applied, and if so, another flag is coded to indicate the direction. In intra coding, RDPCM is applied only when intra prediction mode is either horizontal or vertical and no flag is coded since the RDPCM direction is inferred from the intra prediction mode.



\subsection{Methods based on pixel-by-pixel spatial prediction}
The second group of methods can be used only in lossless intra coding and are based on replacing the block-based spatial prediction method with a pixel-by-pixel spatial prediction method. 
Since the transform is skipped in lossless coding, a pixel-by-pixel spatial prediction approach can be used instead of block-based prediction for more efficient prediction.

The literature contains many lossless intra coding methods based on the pixel-by-pixel prediction approach \cite{SAP,DC2,Templatebased}. The so-called Sample-based Angular Prediction (SAP) method is a well-known such method \cite{SAP}. In the application of the SAP method to HEVC \cite{SAP}, only the angular intra modes are modified and the DC and planar intra modes remain unmodified. In these modified angular intra modes, the same angular projection directions and linear interpolation equations of HEVC's intra prediction are used, but the used reference samples are modified. Instead of the the block neighbor pixels, the immediate neighbor pixels are used as reference pixels for prediction, resulting in a pixel-by-pixel prediction version of HEVC's block-based intra prediction. 

Instead of using the HEVC intra prediction equations for pixel-by-pixel spatial prediction, a more general pixel-by-pixel spatial prediction method based on using 3 neighboring pixels in each intra mode of HEVC is developed in \cite{saeed}, and the results report one of the best lossless intra coding performances within HEVC.

While the lossless intra coding methods based on pixel-by-pixel spatial prediction can provide competitive compression performance, their distinguishing feature can also be a drawback. Their pixel-based nature is not congruent with the block-based architecture of video coding standards and introduces undesired pixel-based dependencies in the prediction architecture that can reduce throughput in the processing pipeline of video  encoders and decoders \cite{SAP,saeed}.



\subsection{Methods based on modified entropy coding}
\label{ssec:entropy}
The third group of methods considers entropy coding. In lossy coding, transform coefficients of prediction residuals are entropy coded, while in lossless coding, the prediction residuals are entropy coded. Considering the difference of the statistics of quantized transform coefficients and prediction residuals, several modifications in entropy coding were proposed for lossless coding \cite{kim2010efficient,CABAC3,CABAC1}. The HEVC version 2 includes reversing the scan order of coefficients, using a dedicated context model for the significance map and other tools \cite{HEVCv2,N0044}.

\section{Integer-to-integer (i2i) transforms}
\label{sec:i2i}
Integer-to-integer (i2i) transforms map integer inputs to integer outputs and are invertible \cite{goyal_i2i}. Note that unlike the integer transforms in HEVC \cite{HEVCtr}, which also map integer residual pixels to integer transform coefficients by implementing the transform operations with fixed-point arithmetic, i2i transforms considered here do not increase the dynamic range at the output. Therefore they can be easily used in lossless compression. 

One possible method to obtain an i2i transform is to decompose a known orthogonal  transform into a cascade of plane rotations, and then approximate each plane rotation with a lifting structure \cite{goyal_i2i, sweldens1996lifting}, which can map integer inputs to integer outputs.

\subsection{Plane rotations and the lifting scheme}
\label{ssec:rotlif}
\begin{figure}[tb]
\begin{center}
\begin{minipage}{0.40\linewidth}
\centering
\includegraphics[trim=53 680 450 55,clip,width=\linewidth]{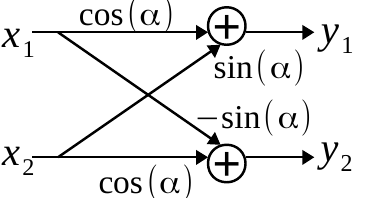}
\centerline{{\footnotesize (a) Plane rotation}}
\end{minipage}
\\ \vspace{0.5cm}
\begin{minipage}{0.54\linewidth}
\includegraphics[trim=53 680 440 55,clip,width=0.81\linewidth]{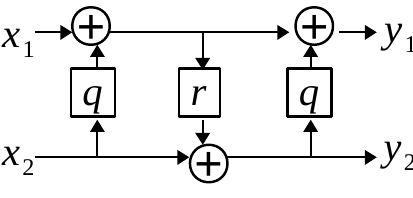}
\centerline{{\footnotesize (b) Decomposition with three lifting steps }}
\end{minipage}
\hfill
\begin{minipage}{0.44\linewidth}
\centering
\includegraphics[trim=53 680 440 55,clip,width=\linewidth]{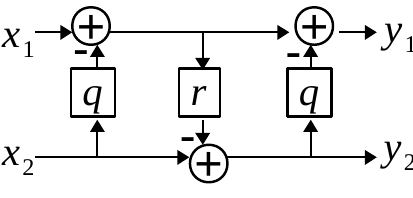}
\centerline{{\footnotesize (c) Inverse structure }}
\end{minipage}
\end{center}
\caption{(a) Plane rotation and (b) its decomposition into a structure with three lifting steps and (c) the inverse structure.}
\label{fig:rot_3lif}
\end{figure}

A plane rotation can be represented with the 2x2 matrix given below in Equation (\ref{eq:pr}) and also shown with a flow-graph in Figure \ref{fig:rot_3lif} (a). 
\begin{align}
\label{eq:pr}
P(\alpha)=
&\left[   \begin{array}{ c c }
     \cos(\alpha) &  \sin(\alpha) \\
    -\sin(\alpha) &  \cos(\alpha) 
  \end{array} \right]
\end{align}
The significance of plane rotations comes from the capability to design orthogonal transforms by cascading multiple plane rotations.

A plane rotation can be decomposed into a structure with three lifting steps or a structure with two lifting steps and two scaling factors \cite{binDCT}. Consider first the decomposition into a structure with three lifting steps as shown in Figure \ref{fig:rot_3lif} (b), which is represented in matrix form as
\begin{align}
\label{eq:rot2}
&\left[   \begin{array}{ c c }
    \cos(\alpha) &  \sin(\alpha) \\
    -\sin(\alpha) &  \cos(\alpha) 
  \end{array} \right]
= &\left[   \begin{array}{ c c }
     1 & q \\
     0 & 1 
  \end{array} \right]
\left[   \begin{array}{ c c }
     1 & 0 \\
     r & 1 
  \end{array} \right]  
\left[   \begin{array}{ c c }
     1 & q \\
     0 & 1 
  \end{array} \right] 
\end{align}
where $q=\frac{\cos(\alpha)-1}{\sin(\alpha)}$ and $r=\sin(\alpha)$.

Each lifting step can be inverted with another lifting step because 
\begin{equation}
\left[   \begin{array}{ c c }
     1 & q \\
     0 & 1 
  \end{array} \right]^{-1}
= \left[   \begin{array}{ c c }
     1 & -q \\
     0 & 1 
  \end{array} \right]  
,
\left[   \begin{array}{ c c }
     1 & 0 \\
     r & 1 
  \end{array} \right]^{-1}
=
\left[   \begin{array}{ c c }
     1 & 0 \\
     -r & 1 
  \end{array} \right].   
\label{eq:lif}  
\end{equation}
In other words, each lifting step is inverted by subtracting out what was added in the forward lifting step. Thus, the inverse of the decomposition structure with 3 lifting steps is obtained by cascading the same lifting steps with subtraction instead of addition in reverse order, as shown in Figure \ref{fig:rot_3lif} (c).

\begin{figure}[tb]
\begin{center}
\begin{minipage}{0.40\linewidth}
\centering
\includegraphics[trim=53 680 450 55,clip,width=\linewidth]{rot_straight.pdf}
\centerline{{\footnotesize (a) Plane rotation}}
\end{minipage}
\\ \vspace{0.5cm}
\begin{minipage}{0.49\linewidth}
\includegraphics[trim=53 680 440 55,clip,width=0.9\linewidth]{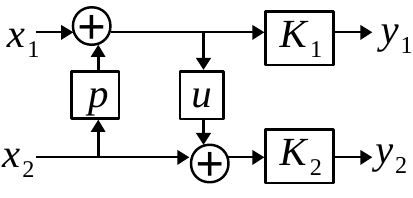}
\centerline{{\footnotesize (b) Type-1 decomposition with 2}} \centerline{{\footnotesize lifting steps and 2 scaling factors}}
\end{minipage}
\hfill
\begin{minipage}{0.49\linewidth}
\centering
\includegraphics[trim=53 680 440 55,clip,width=0.9\linewidth]{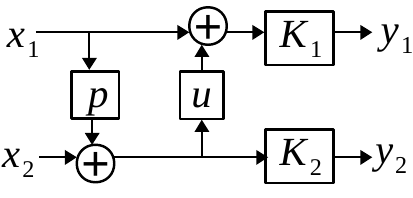}
\centerline{{\footnotesize (c) Type-2 decomposition with 2}} \centerline{{\footnotesize lifting steps and 2 scaling factors}}
\end{minipage}
\\ \vspace{0.5cm}
\begin{minipage}{0.49\linewidth}
\includegraphics[trim=53 680 440 55,clip,width=0.9\linewidth]{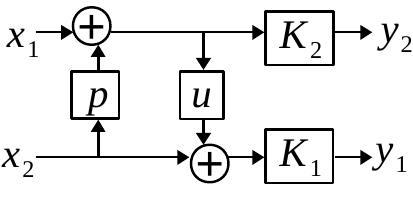}
\centerline{{\footnotesize (d) Type-3 decomposition with 2}} \centerline{{\footnotesize lifting steps and 2 scaling factors}}
\end{minipage}
\hfill
\begin{minipage}{0.49\linewidth}
\centering
\includegraphics[trim=53 680 440 55,clip,width=0.9\linewidth]{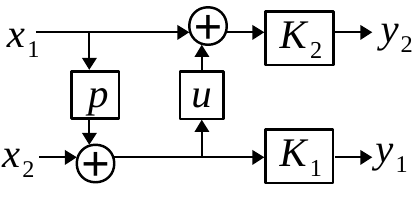}
\centerline{{\footnotesize (e) Type-4 decomposition with 2}} \centerline{{\footnotesize lifting steps and 2 scaling factors}}
\end{minipage}
\end{center}
\caption{(a) Plane rotation and its decomposition into structures with two lifting steps and two scaling factors. There are four possible decompositions as shown in (b), (c), (d) and (e). The decomposition in (d) and (e) have permuted outputs.}
\label{fig:rot_2lif}
\end{figure}

Consider now the decomposition of a plane rotation into a structure with two lifting steps and two scaling factors. There are four such possible decompositions, as shown in Figure \ref{fig:rot_2lif}. Note that the type-3 and type-4 decompositions in Figure \ref{fig:rot_2lif} (d) and (e) have permuted outputs. In other words, output $y_2$ (and scaling factor $K_2$) is now in the upper branch and output $y_1$ (and scaling factor $K_1$) in the lower.

These decompositions can also be represented in matrix form. For example, the decomposition in Figure \ref{fig:rot_2lif} (b) can be represented as in Equation (\ref{eq:lif2type1}) below. 
\begin{align}
\label{eq:lif2type1}
&\left[   \begin{array}{ c c }
    \cos(\alpha) &  \sin(\alpha) \\
    -\sin(\alpha) &  \cos(\alpha) 
  \end{array} \right]
=& \left[   \begin{array}{ c c }
     K_1 & 0 \\
     0 & K_2 
  \end{array} \right] 
\left[   \begin{array}{ c c }
     1 & 0 \\
     u & 1 
  \end{array} \right]  
\left[   \begin{array}{ c c }
     1 & p \\
     0 & 1 
  \end{array} \right]  
\end{align}
%

The lifting parameters $p$ and $u$ and the scaling factors $K_1$ and $K_2$ in all four types of decompositions can be related to the rotation angle $\alpha$ of the plane rotation by first writing the linear equations relating the inputs to the outputs for the decompositions and the plane rotation and then equalizing the linear equations. This results in the following relations.

For the type-1 decomposition in Figure \ref{fig:rot_2lif} (b), the lifting and scaling parameters are related to rotation angle $\alpha$ as follows :
\begin{itemize}
  \item $p=\tan(\alpha)$,~ $u=-\sin(\alpha)\cos(\alpha)$ 
  \item $K_1=\cos(\alpha)$,~ $K_2=\frac{1}{\cos(\alpha)}$.
\end{itemize}

For the type-2 decomposition in Figure \ref{fig:rot_2lif} (c), 
the relations are as follows :
\begin{itemize}
  \item $p=-\tan(\alpha)$,~ $u=\sin(\alpha)\cos(\alpha)$ 
  \item $K_1=\frac{1}{\cos(\alpha)}$,~ $K_2=\cos(\alpha)$.
\end{itemize}

For the type-3 decomposition in Figure \ref{fig:rot_2lif} (d), 
the relations are as follows :
\begin{itemize}
  \item $p=-\frac{1}{\tan(\alpha)}$,~ $u=\sin(\alpha)\cos(\alpha)$ 
  \item $K_2=-\sin(\alpha)$,~ $K_1=\frac{1}{\sin(\alpha)}$.
\end{itemize}

Finally, for the type-4 decomposition in Figure \ref{fig:rot_2lif} (e), the lifting and scaling parameters are related to rotation angle $\alpha$ as follows :
\begin{itemize}
  \item $p=\frac{1}{\tan(\alpha)}$,~ $u=-\sin(\alpha)\cos(\alpha)$ 
  \item $K_2=-\frac{1}{\sin(\alpha)}$,~ $K_1=\sin(\alpha)$.
\end{itemize}

Note that all for types of decomposition structures in Figure \ref{fig:rot_2lif} are equivalent with the above parameters, i.e. they have the same input-output relation. 



Note also that all four types of decompositions are equivalent to the plane rotation in Figure \ref{fig:rot_2lif} (a), i.e. they have the same input-output relation, except that type-3 and type-4 decompositions have permuted outputs, which is just a simple reordering of the output signal. However, when designing i2i transforms, the lifting parameters $p$ and $u$ can be quantized and the scaling factors $K_1$ and $K_2$ can become important, and therefore one type of decomposition can be preferred over the others despite all having the same input-output relation. This issue will be discussed in more detail in Section \ref{ssec:i2idst} where we discuss the design of i2i approximation of the odd type-3 DST (ODST-3) based on lifting decompositions of cascaded plane rotations.

Inversion of decompositions with two lifting steps and two scaling factors can be achieved by going in the reverse direction and inverting first the scaling factors and then the lifting steps. 


\subsection{Integer-to-integer mapping property}
Consider now the integer-to-integer mapping property of the lifting steps. In all of the above decompositions, each lifting step can map integers to integers by introducing a simple rounding operation. If the result of multiplying integer input samples with lifting paramters $p$ or $u$ is rounded to integers, each lifting step performs mapping from integer inputs to integer outputs \cite{goyal_i2i,binDCT}. Notice that as long as the same rounding operation is applied in both forward and inverse lifting steps, inversion of a lifting step remains the same, i.e. subtract what was added in the forward lifting step. In summary, each lifting step can map integers to integers (and is still easily inverted) by introducing rounding operations after multiplications with lifting parameters $p$ or $u$.

The scaling factors in the decompositions in Figure \ref{fig:rot_2lif} violate integer-to-integer mapping property if scaling factors are not integers. If they are integers, they just introduce artificial scaling that is unnecessary. Thus scaling factors seem to pose a problem for integer-to-integer mapping property of the lifting decompositions in Figure \ref{fig:rot_2lif}, however, we discuss in Section \ref{ssec:i2idst} how to deal with scaling factors when designing i2i transforms from cascaded lifting decompositions.

Floating point multiplications can be avoided in lifting steps if the lifting parameters $p$ and $u$ are approximated with rationals of the form $k/2^l$ ($k$ and $l$ are integers), which can be implemented with only integer addition and bitshift operations (integer multiplications can be performed with addition and bitshift). Note that the bitshift operation implicitly includes a rounding operation, which provides integer-to-integer mapping, as discussed above. Integers $k$ and $l$ can be chosen depending on the desired accuracy to approximate the lifting operation and the desired level of computational complexity. 

\subsection{I2i DCT}
\label{ssec:i2idct}
\begin{figure}[t]
  \begin{center}
    \includegraphics[trim=40 600 280 50,clip,width=0.86\linewidth]{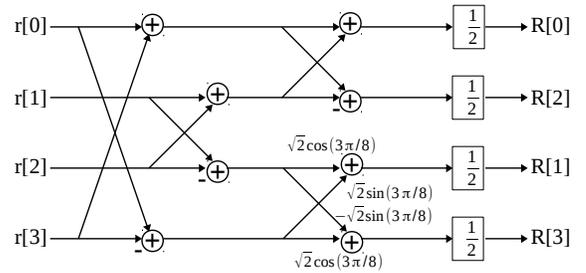}
    \caption{Factorization of 4-point DCT.}
    \label{fig:LoeffFact4}
  \end{center}
\end{figure}

A significant amount of work on i2i transforms has been done to develop i2i approximations of the discrete cosine transform (DCT). One of the most popular methods, due its to lower computational complexity, is to utilize the factorization of the DCT into plane rotations and butterfly structures \cite{Chen,Loeffler,binDCT}. 
Two well-known factorizations of the DCT into plane rotations and butterflies are the Chen's and Loeffler's factorizations \cite{Chen,Loeffler}. Loeffler's 4-point DCT factorization is shown in Figure \ref{fig:LoeffFact4}. It contains three butterflies, one plane rotation and a scaling factor of $\frac{1}{2}$ at the end of each branch. 

Consider first the three butterfly structures shown in Figure \ref{fig:LoeffFact4}. A butterfly structure maps integers to integers because the output samples are the sum and difference of the inputs. It is also easily inverted by itself followed by division of output samples by 2. 

The plane rotation in Figure \ref{fig:LoeffFact4} can be decomposed into three lifting steps or two lifting steps and two scaling factors, as discussed in Section \ref{ssec:rotlif}, to obtain integer-to-integer mapping. Using two lifting steps reduces the complexity and the two scaling factors can be combined with the other scaling factors at the output. 

\begin{figure}[th]
  \begin{center}
    \includegraphics[trim=40 600 300 50,clip,width=0.8\linewidth]{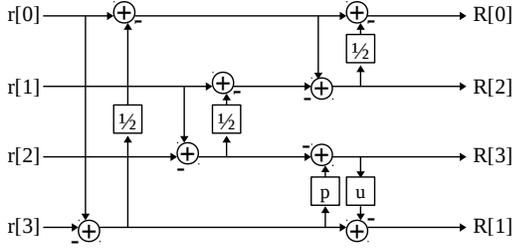}
    \caption{Lifting-based i2i approximation of DCT for lossless compression.}
    \label{fig:binDCT4}
  \end{center}
\end{figure}

The scaling factors at the output can be absorbed into the quantization stage in lossy coding. In lossless coding, all scaling factors can be omitted. However, care is needed when omitting scaling factors since for some branches, the dynamic range of the output may become too high when scaling factors are omitted. For example, in Figure \ref{fig:LoeffFact4}, the DC output sample (i.e. $R[0]$) becomes the sum of all input samples when scaling factors are omitted, however, it may be preferable that it is the average of all input samples, which can improve the entropy coding performance \cite{binDCT}. Hence, to obtain an i2i DCT for use in lossless coding, the butterflies of Figure \ref{fig:LoeffFact4} are replaced with lifting steps to adjust the dynamic range at the output of each branch (or equivalently to adjust the norm of each analysis basis function) and the scaling factors at the output are omitted, resulting in the i2i DCT shown in Figure \ref{fig:binDCT4} \cite{binDCT}.

\section{Integer-to-integer Approximation of Odd Type-3 DST}
\label{sec:i2idst}

To the best of our knowledge, an integer-to-integer (i2i) approximation of the odd type-3 DST (ODST-3) has not appeared in the literature. To develop such an i2i approximation of the ODST-3, we first approximate the ODST-3 with a cascade of plane rotations, and approximate these rotations with lifting steps to obtain i2i approximations of the ODST-3 for use in lossless intra-frame coding.

An overview of this section is as follows. In Section \ref{ssec:acorr}, the auto-correlation expression of the block-based spatial prediction residual and its optimal transform as the correlation coefficient approaches 1, i.e. the ODST-3, are reviewed. Next, in Section \ref{ssec:codg}, a coding gain expression is presented. In Section \ref{ssec:rots}, an algorithm to approximate the 4-point ODST-3 through plane rotations is presented. In Section \ref{ssec:i2idst}, the plane rotation based approximation is used to obtain i2i approximations of the 4-point ODST-3. Finally, in Section \ref{ssec:dst8x8}, i2i approximations of ODST-3 for large block sizes are discussed.

\subsection{Block-based spatial prediction, auto-correlation of its residual and the odd type-3 DST (ODST-3)}
\label{ssec:acorr}
\begin{figure}[t]
  \centering
    \includegraphics[trim=42 582 400 42,clip,width=0.40\linewidth]{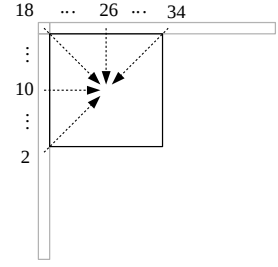}
    \caption{Copying directions of intra prediction modes in HEVC. Modes 2-34 are angular copying modes with the above shown directions and modes 0 and 1 are non-angular DC and planar prediction modes, respectively \cite{IntraHEVC}. }
    \label{fig:ip}
\end{figure}

Block-based spatial prediction, or also commonly called intra prediction, is a widely used technique for predictive coding of intra-frames in modern video coding standards \cite{Luthra264, IntraHEVC}. In this well-known method, a block of pixels are predicted by  copying the block's spatially neighbor pixels (which reside in the previously reconstructed left and upper blocks) along a predefined direction inside the block \cite{IntraHEVC}. While H.264 supports 8 such directional intra prediction modes (each copying spatial neighbors along different directions) in 4x4 and 8x8 blocks, HEVC supports 33 such modes (shown in Figure \ref{fig:ip}) for blocks of sizes 4x4, 8x8, 16x16 and 32x32. The prediction residual block, obtained by subtracting the prediction block from the original block, is transformed and quantized in lossy coding or processed with DPCM in lossless coding in these standards, prior to entropy coding.

The optimal transform for the lossy coding of the spatial prediction residual block was determined as the hybrid DCT/ODST-3 based on modeling the image pixels with a first-order Markov process \cite{Chuo_dst, Han_dst}. Depending on the copying direction of the prediction mode, the DCT or the ODST-3 is applied in either the horizontal and/or vertical direction forming a hybrid 2D transform. In particular, if the copying direction of the prediction mode is horizontal, the ODST-3 is applied along the horizontal direction and the DCT is applied along the vertical direction. Similarly, if the copying direction of the prediction mode is vertical, the ODST-3 is applied along the vertical and the DCT along the horizontal direction. 

Note that although a mode-dependent hybrid transform approach was derived in \cite{Chuo_dst}, compression experiments have shown that using the 2D ODST-3 for all intra modes gives similar compression performance in lossy coding in HEVC, and the HEVC standard uses 2D ODST-3 for all 4x4 intra modes \cite{HEVC}. Based on this result, we also use i2i approximations of 2D ODST-3 for all intra modes in our experiments in Section \ref{sec:expres}. 


Now, we briefly review the derivation of the auto-correlation of the block-based spatial prediction residual because it will be used to develop i2i transforms that approximate the ODST-3 for lossless intra-frame compression. We use a 1D signal in our discussion for simplicity and because the result can be used for 2D signals by constructing separable 2D transforms as in \cite{Chuo_dst, Han_dst}. 

\begin{figure}[t]
  \centering
    \includegraphics[trim=45 700 475 55,clip,width=0.38\linewidth]{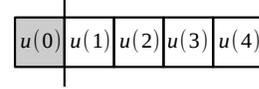}
    \caption{A 4-pixel image row (\textit{white pixels} $u(i)$,~$i=1,..,4$) and its neighbor pixel (\textit{gray pixel} $u(0)$) modeled with a first-order Markov process. The spatial prediction pixels ($\hat{u}(i)$,~$i=1,..,4$) of the block are obtained by copying the block neighbor pixel $u(0)$, in other words, $\hat{u}(i)=u(0)$,~$i=1,..,4$.}
    \label{fig:block}
\end{figure}

A first-order Markov process, which is used to model image pixels horizontally within a row (as shown in Figure \ref{fig:block}) or vertically within a column, is represented recursively as 
\begin{equation}
u(i) = \rho \cdot u(i-1) + w(i)
\end{equation}
where $\rho$ is the correlation coefficient, $u(i)$ are zero-mean, unit variance process samples and $w(i)$ are zero-mean, white noise samples with variance $1-\rho^2$. The auto-covariance or correlation of the process is given by 
\begin{equation}
E[u(i)\cdot u(j)]=\rho^{|i-j|}. 
\label{eq:R}
\end{equation}
It is well known that the Discrete Cosine Transform (DCT) is the optimal transform for the first-order Markov process as its correlation coefficient $\rho$ approaches the value 1 \cite{Ahmed_dctderivation}. 

The spatial prediction block is obtained by copying the neighbor pixel of the block, i.e. $u(0)$, inside the block. In other words, the spatial prediction pixels $\hat{u}(i)=u(0)$,~$i=1,..,N$, where $N$ is the block length.

The residual block pixels $r(i)$,~$i=1,..,N$, are obtained by subtracting the spatial prediction pixels $\hat{u}(i)$ from the original pixels $u(i)$ :
\begin{align}
r(i) &= u(i) - \hat{u}(i) \nonumber \\
     &= u(i) - u(0).
\label{eq:res}
\end{align}
The auto-correlation of the residual pixels is given by $E[r(i)r(j)]$ and is obtained as follows : 
\begin{align}
E[r(i)r(j)] &= E[(u(i) - u(0)) (u(j) - u(0))]  \nonumber \\
            &= \rho^{|i-j|} - \rho^i -\rho^j + 1,  \qquad i,j\in \{1,...,N\}
\label{eq:ac}
\end{align}

Such an auto-correlation expression results in a special auto-correlation matrix as the correlation coefficient $\rho$ approaches 1. In particular, for a block size of $N=4$, the following correlation matrix $K_4$ is obtained :
\begin{equation}
K_4 = 
\begin{bmatrix}
    1 & 1 & 1 & 1  \\
    1 & 2 & 2 & 2  \\
    1 & 2 & 3 & 3  \\
    1 & 2 & 3 & 4  
\end{bmatrix}.
\label{eq:M}
\end{equation}
The eigenvectors of such correlation matrices have been determined to be the basis vectors of the odd type-3 discrete sine transform (ODST-3) given by \cite{Chuo_dst,jain1979sinusoidal, Han_dst} 
\begin{equation}
[S]_{m,n} = \dfrac{2}{\sqrt{2N+1}} ~ sin(\dfrac{(2m-1)n\pi}{2N+1}), \quad m,n\in \{1,...,N\}
\label{eq:odst3}
\end{equation}
where $m$ and $n$ are integers representing the frequency and time index of the basis functions, respectively. Hence, the optimal transform for the spatial prediction residual block is the ODST-3, as $\rho$ approaches 1. 

An important observation regarding the ODST-3 is that its first ($m=1$) and most important basis function has smaller values at the beginning (i.e. closer to the prediction boundary) and larger values towards the end of the block. This trend in the values of the basis function is due to the fact that block pixels closer to the prediction boundary are predicted better than those further away from it, i.e. the variance of the prediction residual signal samples grows with the distance of the samples from the prediction boundary \cite{Chuo_dst,jain1979sinusoidal,Han_dst}.

\subsection{Coding gain in lossy and lossless transform coding}
\label{ssec:codg}

In lossy transform coding, the transform design problem reduces to searching for an orthogonal transform that minimizes the product of the transform coefficient variances \cite{goyal_tc}. The optimal solution, i.e. transform, is given by the eigenvectors of the source correlation matrix, and the most commonly used name for this transform is the Karhunen-Loeve transform (KLT). Based on the transform design problem, a figure of merit called the coding gain $G$ of an orthogonal transform $T$ is defined in the literature as follows :
\begin{equation}
G(T,K_N) = 10 \log_{10} \dfrac{ (\prod_{i=1}^N \sigma_{r,i}^2 )^{\frac{1}{N}} }{ (\prod_{i=1}^N \sigma_{R,i}^2  )^{\frac{1}{N}} }
\label{eq:cg_gen}
\end{equation}
Here, $N$ is the block length of the signal $r(i)$, $i=1,...,N$, $K_N$ is the correlation matrix of the signal with diagonals $\sigma_{r,i}^2$, i.e. $\sigma_{r,i}^2$ is the variance of the $i^{th}$ input sample, $\sigma_{R,i}^2$ is the variance of the $i^{th}$ transform coefficient, i.e. $i^{th}$ output sample. Note that this coding gain expression is obtained under assumptions such as Gaussian source, high-rate quantization and optimal bit allocation \cite{goyal_tc}.

In this paper, we are primarily interested in lossless coding, in particular with integer-to-integer (i2i) transforms. Goyal shows in \cite{goyal_i2i} that under similar assumptions such as Gaussian source and optimal bit allocation, the i2i transform design problem for lossless coding reduces to a similar search for a transform that minimizes, again, the product of the transform coefficient variances, but the search is over all transforms with a determinant of 1 (instead of over orthogonal transforms as in lossy transform coding.) Since we construct i2i transforms in this paper from cascaded lifting steps, all of the i2i transforms in this paper have a determinant of 1 (since each pair of $p$ and $u$ lifting steps has a determinant of 1). Hence, in this paper we use the same coding gain expression in Equation (\ref{eq:cg_gen}) to design and evaluate performances of also i2i transforms to be used for lossless compression.

Notice that the search in the i2i transform design problem is over all transforms with a determinant of 1, instead of over all orthogonal transforms as in transform design for lossy transform coding \cite{goyal_i2i}. Since all orthogonal transforms have a determinant of 1, the search in the i2i transform design is over a larger set of transforms and thus the coding gain obtained with i2i transforms can be larger than that of the KLT, i.e. the maximum obtainable with orthogonal transforms \cite{goyal_i2i}.

In summary, one of the most important metrics of a transform used in compression applications is the coding gain. A transform with higher coding gain 
can achieve higher compression performance (provided following processing stages such as quantization -- if present -- and entropy coding are performed properly.) In this paper, we use the coding gain expression in Equation (\ref{eq:cg_gen}) to design i2i transforms for lossless compression and to evaluate/compare performances of various transforms.

\subsection{Approximation of 4-point odd type-3 DST (ODST-3) through plane rotations}
\label{ssec:rots}
While the widely used DCT has computationally efficient factorizations based on butterfly structured implementations \cite{Chen,Loeffler,binDCT}, such exact factorizations of the odd type-3 DST (ODST-3) do not exist. This is because the denominator $2N+1$ of the ODST-3's basis function in Equation (\ref{eq:odst3}) is not a composite number (i.e. can not be decomposed into product of small integers), in particular, not a power of 2 \cite{btfADST}.

While exact factorizations based on butterflies and plane rotations are not possible for the ODST-3, it is still possible to seek approximations of the transform by  cascading plane rotations. In this section, we discuss a general framework for such approximations and measure the approximation accuracy via the coding gain, defined in Equation (\ref{eq:cg_gen}).

A plane rotation with an angle $\alpha$ that processes the $i^{th}$ and $j^{th}$ branches  of a length $N$ signal can be represented with the following $N$x$N$ matrix :
\newcommand{\Tu}[1]{\ensuremath{T_{U_#1}}}
\begin{equation}
P(i,j,\alpha) =  
       \begin{bmatrix}   1   & \cdots &    0   & \cdots &    0   & \cdots &    0   \\
                      \vdots & \ddots & \vdots &        & \vdots &        & \vdots \\
                         0   & \cdots & \cos\alpha & \cdots & \sin\alpha & \cdots &    0   \\
                      \vdots &        & \vdots & \ddots & \vdots &        & \vdots \\
                         0   & \cdots & -\sin\alpha & \cdots & \cos\alpha  & \cdots &    0   \\
                      \vdots &        & \vdots &        & \vdots & \ddots & \vdots \\
                         0   & \cdots &    0   & \cdots &    0   & \cdots &    1
       \end{bmatrix}
\label{eq:P}       
\end{equation}
where the four sinusoidal terms appear at the intersections of the $i^{th}$ and $j^{th}$ rows and columns. In particular, the non-zero elements of $P(i,j,\alpha)$ are  given by :
\begin{align}
[P(i,j,\alpha)]_{i,i} &= cos\alpha \nonumber \\
[P(i,j,\alpha)]_{j,j} &= cos\alpha \nonumber \\
[P(i,j,\alpha)]_{j,i} &= -sin\alpha \nonumber \\
[P(i,j,\alpha)]_{i,j} &= sin\alpha  \nonumber \\
[P(i,j,\alpha)]_{k,k} &= 1,  &  \ k \ne i,\,j .
\end{align}

When cascading plane rotations, the degrees of freedom for each plane rotation $P(i,j,\alpha)$ are the pair of branches ($i$,$j$) to process with the plane rotation, and the rotation angle $\alpha$. Hence, in cascading plane rotations to approximate the ODST-3, the problem reduces to finding a given number $L$ of ordered branch-pairs ($i_k$,$j_k$) and rotation angles $\alpha_k$ so that the coding gain of the cascaded plane rotations, i.e. the obtained overall transform $\Pi_{k=1}^L P(i_k,j_k,\alpha_k)$, is maximized for a block-based spatial prediction residual signal $r(i)$, $i\in \{1,...,N\}$, with correlation matrix $K_N$ whose entries are given by the correlation expression in Equation (\ref{eq:ac}). This problem can be formalized as the following optimization problem :
\begin{align}
\label{eq:opt}
&\max_{i_1,j_1,\alpha_1,...,i_L,j_L,\alpha_L} G_{}(~ \Pi_{k=1}^L P(i_k,j_k,\alpha_k) ~, ~K_N~) \\
&\qquad \text{subject to} \quad i_k\neq j_k, ~\alpha_k\in [0,\pi/2). \nonumber
\end{align}

This optimization problem does not have a simple solution. The optimization parameters $i_k$ and $j_k$, $k \in \{1,...,L\}$ are discrete and each of them takes an integer value from the set $\{1,...,N\}$. Thus the search space for all the discrete optimization parameters ($i_1,j_1,i_2,j_2,...,i_L,j_L$) contains about $\binom{N}{2}^L$ points since there are about $\binom{N}{2}^L$ many ways to choose the $L$ ordered branch-pairs to which cascaded plane rotations can be applied. The optimization function $G$ does not have any special properties over this discrete search space and each point in it must be exhaustively searched. For each search point, i.e. each possible ordered branch-pair, the rotation angles $(\alpha_1,...,\alpha_L)$ need to be searched, too, to find the maximum of the optimization function $G$, i.e. overall coding gain. 
In summary, to find an optimal or near-optimal solution to the optimization problem in Equation (\ref{eq:opt}), one needs to exhaustively search the space of the discrete optimization parameters ($i_1,j_1,i_2,j_2,...,i_L,j_L$), and for each point in the search space, the rotation angles $(\alpha_1,...,\alpha_L)$ can be searched by employing a gradient-descent type algorithm.

As block size $N$ increases, the described solution approach becomes quickly computationally unmanageable. The number of points $\binom{N}{2}^L$ in the search space of the discrete parameters grows quickly with $N$. In particular, assuming a total of $L=\frac{N}{2}\log_2N$ plane rotations (i.e. similar number of rotations as in an N-point FFT \cite{Lim_book}) the total number of search points $\binom{N}{2}^L\simeq (\frac{N^2}{2})^{\frac{N}{2}\log_2N}$. For a block size of $N=4$, this corresponds to about $2^{12}$ search points, which is manageable, however, for a block size of $N=8$, the number of search points becomes about $2^{60}$, which is too large. Hence for block sizes larger than $N=4$, a different approach is required. A possible approach is to use a faster but sub-optimal greedy algorithm, as in \cite{Zeng_spl}, to solve the optimization problem in a stage-by-stage manner. In each stage, only one rotation $P(i_k,j_k,\alpha_k)$ is considered and its coding gain is maximized by using the output signal of the previous stage as the input. However, such a greedy approach provides solutions with significantly lower coding gains than the KLT in our implementation results. An alternative approach is to use the even type-3 DST (EDST-3) \cite{jain1979sinusoidal}, which can be factored into a cascade of plane rotations \cite{wang1984fast}, as an approximation to the ODST-3 \cite{btfADST}. We pursue the latter approach for designing i2i transforms for lossless compression of block-based spatial prediction residuals with large blocks and discuss this topic further in Section \ref{ssec:dst8x8}. In this section, we continue our discussion for a block size of $N=4$.


Hence, for a spatial prediction residual block of size $N=4$ and a correlation parameter of $\rho=0.95$, we solve the optimization problem in Equation (\ref{eq:opt}) with the above described solution approach. 
In particular, we exhaustively search the space of the discrete optimization parameters ($i_1,j_1,i_2,j_2,...,i_L,j_L$), and for each point in the search space, we search for the best rotation angles $(\alpha_1,...,\alpha_L)$ by employing the optimization toolbox of Matlab. We obtain the solutions for different number of total plane rotations $L$. The coding gains calculated from Equation (\ref{eq:cg_gen}) of the resulting approximations, along with other common transforms, are shown in Table \ref{tb:cg1}.

\begin{table}[t]
\setlength{\tabcolsep}{0.35em}
  \caption{Theoretical coding gains (in dB) of various orthogonal transforms relative to that of the KLT, all applied to the block-based spatial prediction residual  with a block size of $N=4$ and correlation parameter $\rho=0.95$.}
  \label{tb:cg1}
  \centering
    \begin{tabular}{ccccccc}
    DCT  &  ODST-3  & AODST-3$^{(2)}$ & AODST-3$^{(3)}$ & AODST-3$^{(4)}$ & AODST-3$^{(5)}$ \\ \hline 
    -0.6211 & -0.0009 & -0.7593 & -0.1023 & -0.0059 & -0.0001
    \end{tabular}
\end{table}

The results in Table \ref{tb:cg1} are given in terms of coding gain relative to that of the optimal transform, KLT, which achieves a coding gain of $10.0039$ dB. The DCT has, as expected, a big coding gain loss of $0.6211$ dB. The ODST-3 has a coding gain loss of only $0.0009$ dB, since it is optimal as $\rho$ approaches 1. The remaining transforms AODST-3$^{(L)}$ in Table \ref{tb:cg1} represent the obtained approximations to the ODST-3 with $L$ cascaded plane rotations. Their coding gain losses are $0.7593$ dB with 2 cascaded plane rotations, and drop to $0.1023$ dB and $0.0059$ dB with 3 and 4 cascaded plane rotations, respectively. With 5 plane rotations, the coding gain loss is only $0.0001$ dB.

\begin{figure}[t]
  \centering
    \includegraphics[trim=45 590 295 50,clip,width=0.80\linewidth]{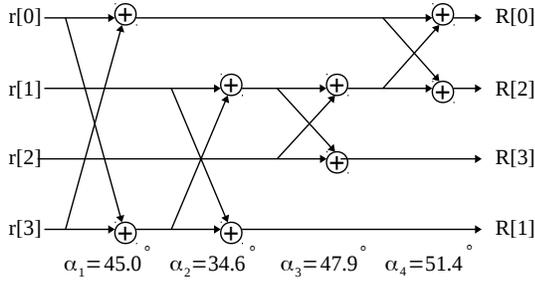}
    \caption{AODST-3$^{(4)}$, the obtained cascade of 4 plane rotations to approximate the 4-point ODST-3. The output branches are labeled according to their variances, i.e. $R[0]$ has the largest variance and $R[3]$ the smallest.}
    \label{fig:aodst3-4}
\end{figure}

The approximation with four cascaded plane rotations, AODST-3$^{(4)}$, is shown in Figure \ref{fig:aodst3-4} with the branch pairs and rotation angles of each plane  rotation. The output branches are labeled according to their variances, i.e. $R[0]$ has the largest variance and $R[3]$ the smallest. AODST-3$^{(4)}$ has a very small coding gain loss relative to the KLT and also uses the same number of rotations as the factorization of DCT in Figure \ref{fig:LoeffFact4}. Hence, we focus on AODST-3$^{(4)}$ in the next section to design i2i approximations of the 4-point ODST-3. 

Note that the coding gains for the AODST-3$^{(L)}$ we have listed in Table \ref{tb:cg1} are the best coding gains we obtained from our optimization problem using our described solution approach. However, we observed from our solution approach that there are also other near-optimal solutions, i.e. cascaded plane rotations that have very close coding gains to the ones in Table \ref{tb:cg1}.

Note also that the obtained approximations with smaller number of rotations are not necessarily prefixes of the ones with more rotations. For example, AODST-3$^{(3)}$ is not equivalent to the cascade of the first three plane rotations in Figure \ref{fig:aodst3-4}. In particular, AODST-3$^{(3)}$ has both different branch-pairs and rotation angles than the first three rotations in Figure \ref{fig:aodst3-4}.

Finally, note that the plane rotations in the obtained AODST-3$^{(L)}$ can, in general, not be applied in parallel unlike in the DCT factorization in Figure \ref{fig:LoeffFact4}, where the first two and last two rotations can be performed in parallel. Of course, our solution to the optimization problem can be modified so that only ordered branch pairs that can be implemented in parallel are used in the search. In this case, the best transform with a total of $L=4$ rotations becomes the one with ordered branch-pairs of (2,4), (1,3), (3,4) and (1,2), and achieves a coding gain of -$0.1206$ dB relative to the KLT.

\subsection{I2i approximation of 4-point odd type-3 DST (ODST-3)}
\label{ssec:i2idst}
This section discusses the design of integer-to-integer (i2i) transforms that approximate the 4-point odd type-3 DST (ODST-3) based on the approximations AODST-3$^{(L)}$ we obtained in the previous section. Although the design approach is general and can be applied to any transform obtained from cascaded plane rotations, we focus on the AODST-3$^{(L)}$ and provide examples based on AODST-3$^{(4)}$. An overview of the remainder of this section is as follows. We first provide a summary of our design approach. Then we discuss how the design approach was developed. Finally, we provide theoretical coding gains for lossless compression with the obtained i2i approximations of the ODST-3.

\begin{figure*}
  \centering
    \includegraphics[trim=00 590 00 50,clip,width=1.00\linewidth]{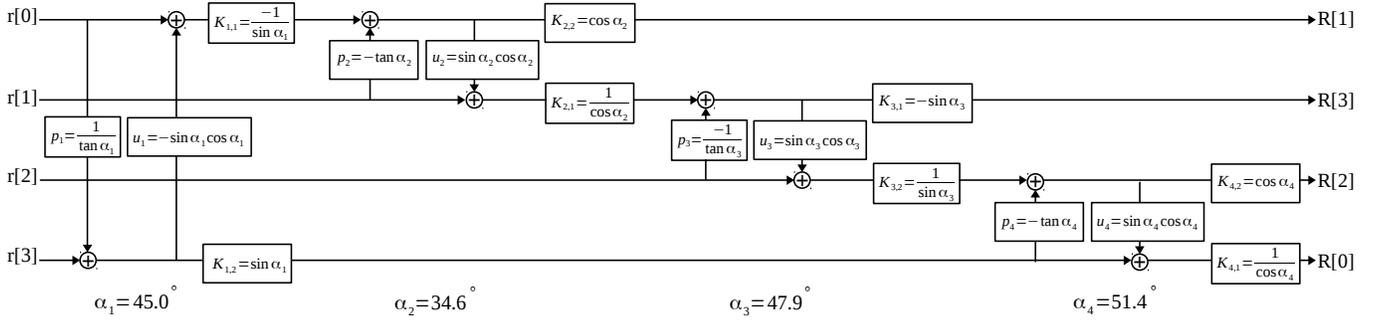}
    \caption{Each plane rotation of the AODST-3$^{(4)}$ in Figure \ref{fig:aodst3-4} is replaced with one of four types of decompositions into two lifting steps and two scaling factors. When a type-3 or type-4 decomposition is used, then the output signal of that decomposition is permuted, which needs to be taken into account for the branches to pair in the following decompositions. The used types for each plane rotation are type-4, type-2, type-3 and type-2, respectively, in this figure.}
    \label{fig:aodst3-lifImp}
\end{figure*}

\begin{figure}
  \centering
    \includegraphics[trim=37 675 260 45,clip,width=1.00\linewidth]{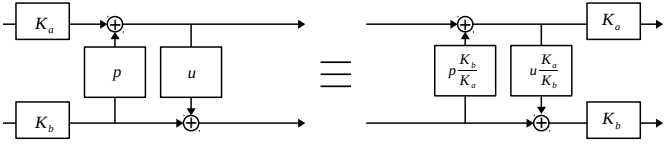}
    \caption{The order of scaling factors $K_a$, $K_b$ and a following lifting structure can be changed so that the scaling factors $K_a$, $K_b$ follow a lifting structure with modified lifting parameters. Note that the above two structures are end-to-end equivalent, i.e. they have the same input-output relation.}
    \label{fig:push}
\end{figure}

\subsubsection{Summary of design approach} \label{sssec:sum} 
\begin{figure*}
  \centering
    \includegraphics[trim=00 605 00 50,clip,width=1.00\linewidth]{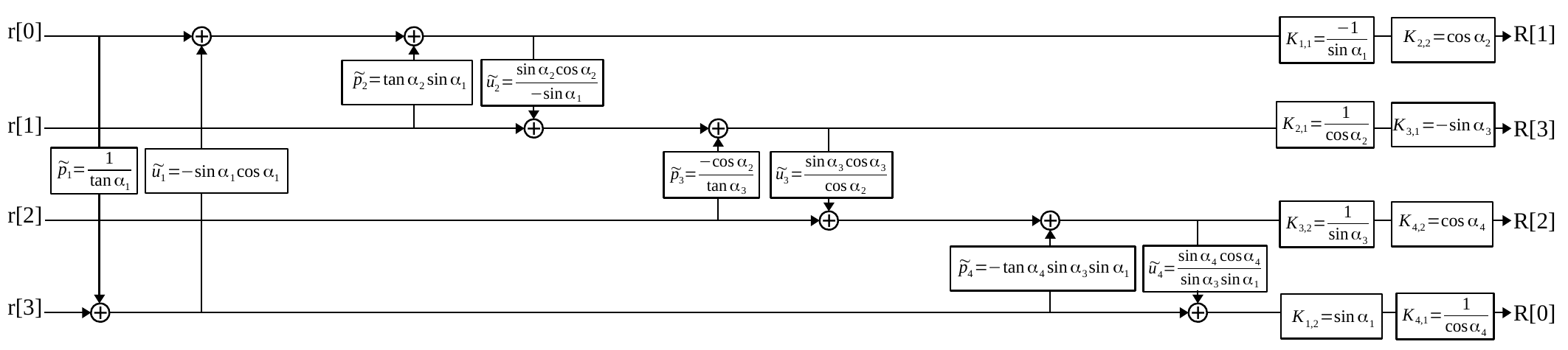}
    \caption{The transform structure obtained after all scaling factors $K_{m,1}$ and $K_{m,2}$ in Figure \ref{fig:aodst3-lifImp} are commuted with following lifting structures so that all scaling factors are pushed to the end of each signal branch.  Note that this new transform structure has the same input-output relation as the transform structure in Figure \ref{fig:aodst3-lifImp} or the AODST-3$^{(4)}$ in Figure \ref{fig:aodst3-4}.}
    \label{fig:aodst3-pushed}
\end{figure*}

\begin{figure}
  \centering
    \includegraphics[trim=0 607 265 50,clip,width=1.00\linewidth]{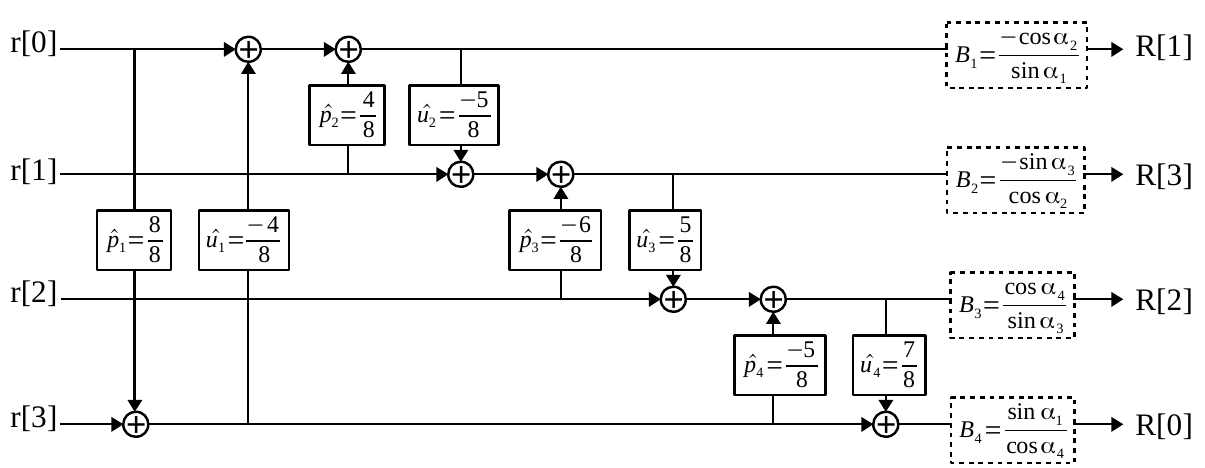}
    \caption{An i2i approximation of the odd type-3 DST (ODST-3) consisting of four cascaded lifting structures with quantized lifting parameters $\hat p$ and $\hat u$.  The following scaling factors $B_i$ can be absorbed into the quantization stage in lossy coding, and omitted in lossless coding. The quantized lifting parameters $\hat p_m$ and $\hat u_m$ are rationals of the form $k/2^l$ ($k$ and $l$ are integers) so that multiplications with them and following rounding operations can be implemented with only integer addition and bit-shift operations. }
    \label{fig:aodst3-i2i}
\end{figure}

Our approach to designing i2i approximations of the ODST-3 can be summarized in 3 steps as follows. 
\begin{enumerate}
  \item[Step] 1 : Given any AODST-3$^{(L)}$, first, each plane rotation is replaced with one of four possible decompositions into two lifting steps and two scaling factors (shown in Figure \ref{fig:rot_2lif}). For AODST-3$^{(4)}$ in Figure \ref{fig:aodst3-4}, a possible result is shown in Figure \ref{fig:aodst3-lifImp}\footnote{Note that the second and fourth rotations of AODST-3$^{(4)}$ in Figure \ref{fig:aodst3-4} connect branches 2 to 4 and 1 to 2, respectively. However, the lifting decompositions of the second and fourth rotations in Figure \ref{fig:aodst3-lifImp} connect branches 2 to 1 and 4 to 3, respectively. This discrepancy comes from using a type-3 or type-4 decomposition, which have permuted outputs, for one or more preceding plane rotations. This issue is discussed in more detail in sub-section \ref{sssec:dev}.}. 
  \item[Step] 2 : Next, the scaling parameters $K_{ 1}$ and $K_{ 2}$ of each decomposition are commuted with the lifting structures of the following decompositions so that all scaling factors are pushed to the end of each signal branch. (This commutative property is discussed in Figure \ref{fig:push}.) For AODST-3$^{(4)}$ with the lifting decompositions in Figure \ref{fig:aodst3-lifImp}, the result of this step is shown in Figure \ref{fig:aodst3-pushed}. 
  \item[Step] 3 : Finally, multiple scaling factors $K_{m,n}$ at the end of each signal branch are combined into one scaling factor $B_i$ per branch, and the updated parameters $\tilde{p}$ and $\tilde{u}$ of the lifting structures are quantized for approximation with rationals of the form $k/2^l$ ($k$ and $l$ are integers) so that multiplications with them and following rounding operations can be implemented with only integer addition and bit-shift operations. A possible result of this step applied to Figure \ref{fig:aodst3-pushed} is shown in Figure \ref{fig:aodst3-i2i}. 
\end{enumerate}

The resulting i2i approximation of the 4-point ODST-3 in Figure \ref{fig:aodst3-i2i} consists of several cascaded lifting structures, with quantized lifting parameters $\hat{p}_m$ and $\hat{u}_m$ followed by scaling factors $B_i$ at the end of each branch. The cascaded lifting structures provide an i2i transform and the scaling factors $B_i$ at the end can be absorbed into the quantization stage in lossy coding, and omitted in lossless coding. 


\subsubsection{Development of the design approach} 
\label{sssec:dev}
Our 3-step design approach described above was developed using the following observations.


Consider first the following observations regarding Step 1.
A plane rotation has equivalent input-output relation with all four types of lifting decompositions into two lifting steps and two scaling factors (see Figure \ref{fig:rot_2lif}), as discussed in Section \ref{ssec:rotlif}. This implies that a plane rotation can be replaced with any of the four types of lifting decompositions. However, when the lifting parameters are quantized, then the input-output relation of the decompositions deviates from that of the plane rotation, and each type of decomposition may incur different quantization error and different deviation. In addition, although all four types of decompositions have equivalent input-output relation, their scaling factors $K_1$, $K_2$ are different, which can become important after steps 2 and 3 are performed, as discussed in sub-section \ref{sssec:typePr}. Thus, the type of decomposition used for each rotation is important and sub-section \ref{sssec:typePr} discusses how to choose the type for each rotation.

Note that type-3 and type-4 lifting decompositions have permuted outputs (see Figure \ref{fig:rot_2lif}), which means that when a type-3 or type-4 lifting decomposition is used to replace a plane rotation, then the output signals in the lower and upper branches are swapped. Hence, when replacing plane rotations in an AODST-3$^{(L)}$ (e.g. Figure \ref{fig:aodst3-4}), the swapping of signals in branches has to be kept track of so that the correct branches are connected in the following lifting decompositions. For example, the first plane rotation in Figure \ref{fig:aodst3-4} is replaced with a type-4 lifting decomposition in Figure \ref{fig:aodst3-lifImp}, which means that the output signal of the first plane rotation in the top (bottom) branch in Figure \ref{fig:aodst3-4} is in the bottom (top) branch in Figure \ref{fig:aodst3-lifImp}. Hence, although the second plane rotation in Figure \ref{fig:aodst3-4} connects branches 2 to 4, the second lifting decomposition in Figure \ref{fig:aodst3-lifImp} must now connect branches 2 to 1. In summary, after each lifting decomposition, one must keep track of which branch in this new transform structure (e.g. Figure \ref{fig:aodst3-lifImp}) contains which branch from the AODST-3$^{(L)}$ structure (e.g. Figure \ref{fig:aodst3-4}), and connect the branches of the following lifting decompositions accordingly.


Now, consider the following observations regarding Step 2.
The order of scaling factors and a following lifting structure can be changed so that the scaling factors follow a lifting structure with modified parameters. This change of order does not change the input-output relation of this local structure and is discussed in Figure \ref{fig:push}. Applying this reordering repeatedly to all scaling factors in an AODST-3$^{(L)}$ implementation with lifting decompositions results in a new transform structure that has the same overall input-output relation but all lifting steps are at the beginning and all scaling factors are at the end of the new overall transform structure. For the AODST-3$^{(4)}$ implementation in Figure \ref{fig:aodst3-lifImp}, the repeated reordering gives the new transform structure in Figure \ref{fig:aodst3-pushed}. It can be verified that this new transform structure in Figure \ref{fig:aodst3-pushed} has the same input-output relation as the transform structure in Figure \ref{fig:aodst3-lifImp} or the AODST-3$^{(4)}$ in Figure \ref{fig:aodst3-4}. 

Finally, consider the following observations regarding Step 3. The lifting parameters $\tilde{p}_m$ and $\tilde{u}_m$ are quantized for approximation with rationals of the form $k/2^l$ ($k$ and $l$ are integers) so that multiplications with them and following rounding operations can be implemented with only integer addition and bit-shift operations, which is a desirable property in video compression. Quantization of lifting parameters also means that the overall transform starts deviating from the AODST-3$^{(L)}$ and the coding gain tends to drop. The choice of $l$ provides a trade-off between approximation accuracy (i.e. coding gain loss) and implementation complexity. A possible result of this step applied to Figure \ref{fig:aodst3-pushed} is shown in Figure \ref{fig:aodst3-i2i}, where $l=3$.

Note that such a transform structure where all lifting steps are at the beginning and all scaling factors are at the end is convenient for obtaining an i2i transform since the lifting steps at the beginning can provide integer-to-integer mapping (as discussed in Section \ref{sec:i2i}), and the scaling factors $B_i$ at the end of each branch can be absorbed into the quantization stage in lossy coding, and omitted in lossless coding.

In lossy coding, the scaling factors $B_i$ can be absorbed into the quantization stage, and this does not change the coding gain of the overall system \cite{malvar2003low, goyal_tc}. 

In lossless coding, when the scaling factors $B_i$ are omitted (i.e. replaced with 1) the coding gain for lossless compression with the i2i transform, again, does not change and can be explained as follows. The denominator of the coding gain expression in Equation (\ref{eq:cg_gen}) becomes $(\prod_{i=1}^N \sigma_{R,i}^2 B_{i}^{-2} )^{\frac{1}{N}}$ when the scaling factors are omitted, where $B_{i}$ are the aggregate scaling factors at the end of the $i^{th}$ branch. Note that $B_{i}$ are obtained by products of several scaling factors $K_{m,n}$ (where $m=1,2,...,L$ represents the plane rotation number and $n=1,2$ indicates either of two scaling coefficients of the decomposition) 
and the product $\prod_{i=1}^N B_i = \prod_{m=1}^L K_{m,1}K_{m,2}$. Since $K_{m,1}=\frac{\pm 1}{K_{m,2}}$ (see Figure \ref{fig:rot_2lif}), the product $\prod_{i=1}^N B_i = \pm 1$ and hence the coding gain does not change when the scaling factors $B_i$ are omitted. 

In summary, given an AODST-3$^{(L)}$, all plane rotations are replaced with one of four types of lifting  decompositions into two lifting steps and two scaling factors. The scaling factors from all lifting decompositions are pushed to the end of each branch using the equality in Figure \ref{fig:push}. The lifting parameters are quantized to rationals of the form $k/2^l$ so that integer arithmetic can be used for the computations. The cascade of lifting steps at the beginning of the structure provides an i2i transform, and the scaling factors at the end of each branch can be absorbed into the quantization stage in lossy coding, and omitted in lossless coding, which do not change the coding gains in lossy or lossless coding.


\subsubsection{Choice of lifting decomposition type for each plane rotation} \label{sssec:typePr}
One issue that was not addressed yet in our i2i transform design approach is which one of the four types of lifting decompositions should be used to replace each plane rotation in a given AODST-3$^{(L)}$. 
Although all four types of lifting decompositions have the same input-output relation as the plane rotation, there are two reasons why one type of decomposition may be preferred over the others to replace a particular plane rotation.

The first reason comes from the quantization of the lifting parameters $\tilde{p}_m$ and $\tilde{u}_m$ in Step 3 of our design approach. When the lifting parameters $\tilde{p}_m$ and $\tilde{u}_m$ are quantized, then each different type of decomposition may incur different quantization error for a particular plane rotation with angle $\alpha_m$ and can affect the overall transform differently. 

The second reason comes from the obtained scaling factors $B_i$ at the end of each branch in Step 3 of our design approach. When scaling factors $B_i$ are omitted in lossless coding, the obtained i2i transform is a scaled AODST-3$^{(L)}$ because with the scaling factors $B_i$, it has the same input-output relation as the AODST-3$^{(L)}$. Then the obtained i2i transform by omitting the scaling factors $B_i$ is simply equal to the AODST-3$^{(L)}$ with its $i^{th}$ analysis basis function multiplied by $B_i^{-1}$. While this scaling does not change the theoretical coding gain for lossless compression as discussed in sub-section \ref{sssec:dev}, it can have a significant impact on compression performance if the entropy coder is not aware of this scaling. In our experiments, we use the reference software of HEVC and its standard entropy coder that is designed for the statistics of the orthogonal DCT or ODST-3. In our compression experiments, we have observed that the best compression performance were achieved by i2i approximations of the ODST-3 where all scaling factors $B_i$ were close to $\pm$1, i.e. the scaling was as small as possible so that the i2i transform was as close as possible to being orthogonal. For example, the scaling factors $B_i$ in Figure \ref{fig:aodst3-i2i} are equal to $-1.1644$, $-0.9013$, $0.8400$ and $1.1344$, respectively. This i2i ODST-3 is one of the "least scaled" i2i approximations of AODST-3$^{(4)}$ and provides the best compression performance results in our experiments. 

In summary, the two reasons why the choice of lifting decomposition type is important when replacing plane rotations are the quantization of the lifting parameters $\tilde{p}_m$ and $\tilde{u}_m$, and the omission of the scaling factors $B_i$ in lossless coding that causes a scaled i2i approximation of AODST-3$^{(L)}$. One simple approach to choose the type of lifting decomposition for each plane rotation in a given AODST-3$^{(L)}$ is to go through all possible combinations of decomposition types for all plane rotations in the AODST-3$^{(L)}$ (i.e. there will be a total of $4^L$ combinations), apply Steps 2 and 3 in the design approach, and choose the combination of types that provides the best coding gain and also the "least scaled" i2i transforms, i.e. scaling factors $B_i$ close to 1. This is how we chose the types of decompositions in Figure \ref{fig:aodst3-lifImp} and we used the resulting i2i approximation of the ODST-3 in Figure \ref{fig:aodst3-i2i} in our experimental results.


\subsubsection{Coding gains for lossless compression} \label{sssec:lscg}
We now provide coding gains for lossless compression with the i2i approximations of the 4-point ODST-3 we designed based on the approach discussed so far. 
In particular, the lifting decomposition based representation of the AODST-3$^{(4)}$ in Figure \ref{fig:aodst3-lifImp} and its equivalent form in Figure \ref{fig:aodst3-pushed} provide one of the best transform structures for i2i approximation of the 4-point ODST-3. Using this transform structure, we provide coding gains for lossless compression based on Equation (\ref{eq:cg_gen}) under different quantization levels of the lifting parameters $\tilde{p}_m$ and $\tilde{u}_m$. 
We quantize the lifting parameters to rationals of the form $k/2^l$ ($k$ and $l$ are integers) and provide the obtained coding gains for various values of $l$. The results are given in Table \ref{tb:cgi2i}.

\begin{table}[t]
\setlength{\tabcolsep}{0.35em}
  \caption{Theoretical coding gains (in dB, relative to that of the KLT) for lossless compression with i2i approximations of 4-point ODST-3 with varying levels of quantization of lifting parameters, all applied to the block-based spatial prediction residual with block size of $N=4$ and correlation parameter $\rho=0.95$.}
  \label{tb:cgi2i}
  \centering
    \begin{tabular}{l|cccccccc}
    $l$      & 8      & 7       & 6      & 5       & 4       & 3       & 2      & 1 \\ \hline 
    $G$ & -0.0059 & -0.0060 & -0.0056 & -0.0104 & -0.0165 & -0.0158 & -0.0973 & -1.0565 
    \end{tabular}
\end{table}

Table \ref{tb:cgi2i} provides the coding gains relative to the coding gain of the KLT.  Note that as the quantization step size becomes arbitrarily small, i.e. $l$ grows arbitrarily large, the obtained i2i transform approaches the AODST-3$^{(4)}$. 
Hence, the i2i transform with large $l$ ($l=8$) has the same coding gain loss ($0.0059$ $dB$) as the AODST-3$^{(4)}$ in Table \ref{tb:cg1}. 
As $l$ is reduced, the coding gain losses increase in general since the obtained i2i transforms deviate more significantly from the AODST-3$^{(4)}$. The coding gain loss for $l\geq 3$ is not significant in practice. For $l=2$, the coding gain drop is about $0.0973$ dB and this can be important in practice. Thus, the quantization level $l=3$ seems to be a good trade-off between coding gain loss and complexity of the i2i transform and we choose this $l$ value, for which the quantized lifting parameters are shown in Figure \ref{fig:aodst3-i2i}, for our compression experiments within HEVC in Section \ref{sec:expres}.

Note that the coding gain for lossless compression with the RDPCM approach, discussed in Section \ref{ssec:rdpcm}, can also be calculated from Equation (\ref{eq:cg_gen}). The coding gain of simple DPCM, applied to the (one-dimensional) block-based spatial prediction residual with a block size of $N=4$ and correlation parameter $\rho=0.95$, is only $0.0039$ dB lower than that of the KLT. This coding gain loss is slightly better than that ($0.0158$ dB) of the i2i transform with $l=3$ that we use in our experiments. However, note that the simple DPCM method is used only along the horizontal or vertical direction in the horizontal and vertical intra modes in HEVC or H.264 and is not used in the other angular intra modes. This is because in the other angular intra modes, the residual exhibits 2D directional correlation and a corresponding directional DPCM, designed separately for each angular intra mode, is required to account for the directional 2D correlation. However, the additional compression benefits do not justify the additional complexity increase and HEVC or H.264 do not use such directional DPCM methods. On the other hand, a 2D  i2i transform based on the designed i2i approximations of the ODST-3 can be used for all intra modes and does not need to be redesigned or optimized for every intra mode.

\subsection{I2i transforms for block-based spatial prediction residuals with large block sizes}
\label{ssec:dst8x8}
The approach we used in Section \ref{ssec:rots} to obtain approximations of the odd type-3 DST (ODST-3) by cascading plane rotations works well for small block sizes, such as $N=4$, but becomes computationally unmanageable for block sizes equal to or larger than $N=8$. Hence, a different approach is required to approximate the ODST-3 for large block sizes. 

One possible approach is to use the even type-3 DST (EDST-3) \cite{jain1979sinusoidal}, which can be factored into a cascade of plane rotations \cite{wang1984fast}, as an approximation for the ODST-3 \cite{btfADST}. Han et al. use the EDST-3 in lossy coding within the VP9 codec to transform the block-based spatial prediction residuals of 8x8 blocks and report compression results very close to those with the ODST-3 \cite{btfADST}. 

The basis functions of the EDST-3 are given by 
\begin{equation}
[E]_{m,n} = \sqrt{\dfrac{2}{N}} ~ sin(\dfrac{(2m-1)(2n-1)\pi}{4N}), \quad m,n\in \{1,...,N\}
\label{eq:edst3}
\end{equation}
where $m$ and $n$ are integers representing the frequency and time index of the basis functions, respectively. When the first ($m=1$) and most important basis function is plotted, one can see that, similar to the first basis function of the ODST-3 in Equation (\ref{eq:odst3}), it has smaller values at the beginning (i.e. closer to the prediction boundary) and larger values towards the end of the block, which implies that the EDST-3 may have good coding gain, in particular better than the conventionally used DCT, for block-based spatial prediction residuals.

Table \ref{tb:cg8x8} lists the coding gain losses of ODST-3 and EDST-3 with respect to the KLT for a spatial prediction residual block with correlation coefficient $\rho=0.95$ at various block sizes. It can be seen from the table that the coding gain loss of EDST-3 with respect to the ODST-3 becomes smaller as block size $N$ increases. The coding gain loss of EDST-3 with respect to ODST-3 is $0.2165$ dB for a block size of $N=4$, drops to $0.1352$ dB for a block size of $N=8$, and drops further for larger block sizes. The coding gains of the DCT are also shown in Table \ref{tb:cg8x8} for comparison.

\begin{table}[t]
  \caption{Theoretical coding gains (in dB) of ODST-3, EDST-3 and DCT relative to that op the KLT, all applied to the block-based spatial prediction residual with correlation parameter $\rho=0.95$ and varying block sizes.}
  \label{tb:cg8x8}
  \centering
    \begin{tabular}{l|cccc}
Block size &  4      &  8      &  16     &  32  \\ \hline 
ODST-3 & -0.0009 & -0.0024 & -0.0045 & -0.0072 \\
EDST-3 & -0.2174 & -0.1376 & -0.0797 & -0.0468 \\
DCT    & -0.6211 & -0.5611 & -0.4108 & -0.2640 \\  
    \end{tabular}
\end{table}

For large block sizes, the coding gain loss arising from using the EDST-3 instead of the ODST-3 can be a good trade-off for the reduction in computational complexity, since the ODST-3 must be implemented with a general matrix multiplication (with complexity $\propto N^2$) while the EDST-3 can be implemented with a cascade of plane rotations with complexity $\propto Nlog_2N$.

The coding gains in Tables \ref{tb:cg1} and \ref{tb:cg8x8} indicate that the AODST-3$^{(4)}$ derived in Section \ref{ssec:rots} has better coding gain than the EDST-3 for a block size of $N=4$. 
In particular, their coding gain losses with respect to that of the KLT are $0.0059$ and $0.2174$ dB, respectively. For larger block sizes, the approach we used in Section \ref{ssec:rots} to obtain the AODST-3$^{(4)}$ becomes computationally unmanageable and we use the EDST-3 to approximate the ODST-3. 

As block size increases, the EDST-3 becomes a better approximation of the ODST-3, i.e. the coding gain loss arising from using the EDST-3 instead of the ODST-3 reduces. For example, for $N=8$ and $N=16$, the coding gain losses drop to $0.1352$ and $0.0752$ dB, respectively. While these coding gain losses may still be considered significant in some contexts, they become insignificant in HEVC, which we use for our experimental results, because these block sizes larger than $N=4$ are used rarely in lossless compression in HEVC. 

The block sizes available in HEVC for intra prediction and transform range from 4x4 to 32x32. However in lossless compression, large block sizes such as $N=8$ or $N=16$ (i.e. 8x8 and 16x16 intra prediction blocks) are used much less frequently than the block size of $N=4$ (i.e. 4x4 intra prediction blocks), as we show in the experimental results in Section \ref{sec:expres}. This is because the bitrate of the prediction residual dominates the overall bitrate in lossless compression (i.e. bitrate of side information, such as intra modes, is a very small fraction of the overall bitrate), and to reduce the bitrate of the residual, better prediction is needed, which is best at the smallest available block size, i.e. 4x4 block size. 

Thus in lossless compression within HEVC (or any other codec that has 4x4 block intra prediction and transforms), lossless compression efficiency of the block size of $N=4$ dominates the overall lossless compression efficiency of the system, and the sub-optimal performance at larger block sizes has an insignificant effect on the overall compression results, as we show in Section \ref{sec:expres}. Nevertheless, to demonstrate this insignificant effect, we design an i2i transform based on the EDST-3 for only the block size $N=8$ (i.e. 8x8 intra prediction blocks) and provide experimental results with it in Section \ref{sec:expres}.


The approach we use to design the 8-point i2i transform based on the 8-point EDST-3, in particular its representation as a cascade of plane rotations, is the same as presented in Section \ref{ssec:i2idst}. We apply the same 3-step procedure. First, each plane rotation in the 8-point EDST-3 is replaced with one of four possible decompositions into two lifting steps and two scaling factors. Next, the scaling factors from all decomposition are pushed to the end of each branch. Finally, multiple scaling factors at the end of each branch are combined into a single scaling factor per branch, and all lifting parameters are quantized for approximation with rationals of the form $k/2^l$, where we use $l=8$. We also chose the type of lifting decomposition for each plane rotation so that the overall i2i transform is as close as possible to being orthogonal. The resulting 8-point i2i transform has a coding gain loss of only $0.0001$ dB relative to the 8-point EDST-3.

\section{Experimental Results}
\label{sec:expres}

The i2i approximation of the 4-point ODST-3 in Figure \ref{fig:aodst3-i2i} and the i2i approximation of the 8-point ODST-3 discussed in Section \ref{ssec:dst8x8} are implemented into the HEVC version 2 Range Extensions (RExt) reference software (HM-15.0+RExt-8.1) \cite{HMref} to provide experimental results of these developed i2i transforms for lossless intra-frame compression. Both of these i2i transforms are applied along first the horizontal and then the vertical direction to obtain 4x4 and 8x8 i2i approximations of the 2D ODST-3 for 4x4 and 8x8 intra prediction residual blocks, respectively. These 2D i2i transforms are used in lossless compression to transform 4x4 and 8x8 block intra prediction residuals of both luma and chroma pictures. 

\subsection{Experimental Setup}
To evaluate the performance of the developed i2i transforms, the following systems are derived from the reference software\footnote{We are planing to share the source code of our modified reference software, from which all these systems can be obtained, on github.com.} and compared in terms of lossless intra-frame compression performance and complexity :
\begin{itemize}
  \itemsep0em
  \item HEVCv1
  \item HEVCv2
  \item i2iDST4
  \item i2iDST4+RDPCM
  \item i2iDST4\&8
  \item i2iDST4\&8+RDPCM.
\end{itemize}
The employed processing in each of these systems is summarized in Table \ref{tb:systems} and discussed below.

\begin{table*}[t]
\centering
\caption{Processing of intra prediction residual blocks prior to entropy coding in each system}
\label{tb:systems}
\begin{tabular}{l|ccccccc}
\hline
                       & HEVCv1  & HEVCv2        & i2iDST4        & i2iDST4       & i2iDST4\&8           & i2iDST4\&8      \\ 
                       &         &               &                &  +RDPCM       &                    &  +RDPCM         \\ \hline \hline
 4x4 hor/ver intra     &   -     & hor/ver rdpcm & 4x4 i2i 2D DST & hor/ver rdpcm & 4x4 i2i 2D DST     & hor/ver rdpcm   \\ \hline
 4x4 other intra       &   -     & -             & 4x4 i2i 2D DST & 4x4 i2i 2D DST& 4x4 i2i 2D DST     & 4x4 i2i 2D DST  \\ \hline
 8x8 hor/ver intra     &   -     & hor/ver rdpcm & hor/ver rdpcm  & hor/ver rdpcm & 8x8 i2i 2D DST     & hor/ver rdpcm   \\ \hline
 8x8 other intra       &   -     & -             & -              & -             & 8x8 i2i 2D DST     & 8x8 i2i 2D DST  \\ \hline
larger hor/ver intra   &   -     & hor/ver rdpcm & hor/ver rdpcm  & hor/ver rdpcm & hor/ver rdpcm      & hor/ver rdpcm   \\ \hline
larger other intra     &   -     & -             & -              & -             & -                  & -               \\ \hline
\end{tabular}
\end{table*}

The HEVCv1 system represents HEVC version 1, which just skips transform and quantization and sends the prediction residual block without any further processing to the entropy coder, as discussed in Section \ref{sec:pr}. 

The HEVCv2 system represents HEVC version 2, in which horizontal RDPCM is applied in the horizontal intra mode at all available block sizes from 4x4 to 32x32, and vertical RDPCM is applied in the vertical intra mode at all available block sizes. For all the other 33 intra modes at all available block sizes, the prediction residual is not processed and sent to the entropy coder. 

The remaining systems employ the developed i2i approximations of the ODST-3. In the i2iDST4 system, the RDPCM system of the HEVC reference software is disabled in 4x4 intra prediction blocks and the 4x4 i2i 2D ODST-3 is used in all modes of 4x4 intra prediction residual blocks. In larger blocks, the default HEVCv2 processing, i.e. RDPCM in horizontal and vertical modes, is used.

In the i2iDST4+RDPCM system, the i2i transform and RDPCM methods are combined in 4x4 block intra coding. In other words, in intra coding of 4x4 intra prediction blocks, the RDPCM method of HEVCv2 is used if the intra prediction mode is horizontal or vertical, and the 4x4 i2i 2D ODST-3 is used for other intra prediction modes. 
In larger blocks, the default HEVCv2 processing, i.e. RDPCM in horizontal and vertical modes, is used.

In the i2iDST4\&8 system, the RDPCM system of the HEVC reference software is disabled in 4x4 and 8x8 intra prediction residual blocks and the 4x4 and 8x8 i2i 2D ODST-3 are used in all modes of 4x4 and 8x8 intra prediction residual blocks. In larger residual blocks, such as 16x16 or 32x32 blocks, the default HEVCv2 processing, i.e. RDPCM in horizontal and vertical modes, is used.

Finally, in the i2iDST4\&8+RDPCM system, the i2i transform and RDPCM methods are combined in 4x4 and 8x8 block intra coding. In other words, in intra coding of 4x4 and 8x8 intra prediction blocks, the RDPCM method of HEVCv2 is used if the intra prediction mode is horizontal or vertical, and the 4x4 or 8x8 i2i 2D ODST-3 is used for other intra prediction modes. In larger blocks, the default HEVCv2 processing, i.e. RDPCM in horizontal and vertical modes, is used. 

Table \ref{tb:systems} summarizes the processing in all systems. In all systems, except HEVCv1 system, available RExt tools, such as a dedicated context model for the significance map, Golomb rice parameter adaptation, intra reference smoothing and residual rotation \cite{HEVCv2,N0044}, are used. However, the residual rotation RExt tool is not used with i2i transforms since i2i transforms already compact the residual energy into the lower frequency transform coefficients.

\subsection{Lossless Intra-frame Compression Results}
For the experimental results, the common test conditions in \cite{commontest} are followed, except that only the first 150 frames 
are coded from every sequence due to our limited computational resources. The results are shown in Table \ref{tb:res}, which include average percentage ($\%$) bitrate reductions and encoding/decoding times of all systems with respect to HEVCv1 system for All-Intra-Main encoding settings \cite{commontest}. 

\begin{table}[tb]
\setlength{\tabcolsep}{3pt}
\centering
\caption{Average percentage ($\%$) bitrate reduction and encoding/decoding times of several systems with respect to the HEVCv1 system in lossless intra coding for All-Intra-Main settings.}
\label{tb:res}
\begin{tabular}{l|ccccc}
\hline
          & HEVCv2  & i2iDST4    & i2iDST4    & i2iDST4\&8   & i2iDST4\&8 \\ 
          &         &            & +RDPCM     &              & +RDPCM  \\ \hline \hline
Class A   & 7.2     & 11.7       &12.1        & 12.1         & 12.6    \\ \hline
Class B   & 4.5     & 6.3        & 6.5        &  6.4         &  6.7   \\ \hline
Class C   & 5.3     & 6.3        & 7.0        &  6.3         &  7.1   \\ \hline
Class D   & 7.5     & 8.4        & 9.4        &  8.2         &  9.5   \\ \hline
Class E   & 8.2     & 9.2        &10.5        &  8.8         & 10.5    \\ \hline
Average   & 6.4     & 8.3        & 8.9        &  8.2         &  9.1    \\ \hline
Enc. T. & $94.6\%$  & $99.0\%$   & $99.6\%$   &  $107.2\%$   & $103.1\%$  \\ \hline
Dec. T. & $92.9\%$  & $95.3\%$   & $98.0\%$   &  $95.8\%$    & $97.0\%$  \\ \hline
\end{tabular}
\end{table}

Consider first the results of the HEVCv2, i2iDST4 and i2iDST4+RDPCM systems in Table \ref{tb:res}. Their average (averaged over all sequences in all classes) bitrate savings with respect to HEVCv1 system are $6.4\%$, $8.3\%$ and $8.9\%$, respectively. Notice also from the results in the table that the systems employing the developed 4-point i2i ODST-3, i.e. i2iDST4 and i2iDST4+RDPCM, achieve consistently larger bitrate reductions than HEVCv2 in all classes.

Note also that the i2iDST4+RDPCM system performs better than the i2iDST4 system in all classes. In other words, the results indicate that if RDPCM is used for horizontal and vertical intra modes, and i2i 2D ODST-3 for other intra modes, as in the i2iDST4+RDPCM system, the best lossless compression performance is achieved. This is because the residual in the horizontal and vertical intra modes can be modeled well with separable 2D correlation (with much larger correlation along the prediction direction than the perpendicular direction) \cite{MrkvIntra} and thus the simple horizontal or vertical DPCM is a great fit and can achieve very good compression performance in these modes, as indicated by its good theoretical coding gain discussed in sub-section \ref{sssec:lscg}. In the remaining intra modes, the horizontal or vertical RDPCM method would not work well (see sub-section \ref{sssec:lscg}) but the designed i2i ODST-3 can provide good compression gains.

Consider now also the results of the i2iDST4\&8 and i2iDST4\&8+RDPCM systems in Table \ref{tb:res}. As shown in Table \ref{tb:systems}, these systems use an i2i approximation of ODTS-3 also in 8x8 blocks, in addition to that in 4x4 blocks. The bitrate savings achieved by these systems, however, do not provide significant or consistent increases on top of those provided by the i2iDST4 and i2iDST4+RDPCM systems. In other words, using i2i ODST-3 in 4x4 blocks seems to provide most of the achievable compression gain and using also an i2i approximation of ODST-3 in 8x8 blocks does not seem to provide significant increase in lossless compression performance. This is reminiscent of the similar situation in lossy coding, where HEVC uses the ODST-3 in only 4x4 intra blocks, and the additional compression gains from using the ODST-3 (instead of the conventional DCT) in lossy coding of larger intra blocks is small and does not justify the additional computational complexity burden of the ODST-3 over the DCT \cite{saxena2011mode}.

Finally, consider also the average encoding and decoding times of all systems in Table \ref{tb:res}. They are compared to those of HEVCv1, assuming HEVCv1 system spends 100\% time on encoding and decoding. The HEVCv2, i2iDST4, and i2iDST4+RDPCM systems achieve lower encoding and decoding times than HEVCv1, despite their additional processing of the residuals, mainly due to their lower bitrates which allow the complex entropy coding/decoding to finish faster. The i2iDST4\&8 and i2iDST4\&8+RDPCM systems have longer encoding times than HEVCv1 since the 8-point i2i approximation of ODST-3 requires more computation than the 4-point i2i ODST-3 or the DPCM method, however, the decoding times are shorter than those of HEVCv1 since the 8-point inverse i2i ODST-3 is rarely used at the decoder, as we discuss in section \ref{ssec:stats}.

The results of the i2iDST4\&8 and i2iDST4\&8+RDPCM systems in Table \ref{tb:res} indicate that when i2i ODST-3 is used in 4x4 blocks, using i2i approximation of ODST-3 in also 8x8 blocks does not improve lossless intra frame compression significantly or consistently. To analyze this result further, we perform a new set of experiments. We disable the use of 4x4 intra prediction blocks in all systems so that the smallest block size for intra prediction is 8x8, which allows us to investigate how the systems compare when only the 8-point i2i ODST-3 is available. In other words, in this new set of experiments, the processing of intra prediction residual blocks prior to entropy coding is the same as in Table \ref{tb:systems}, except that the top two rows with 4x4 block processing are not allowed in all systems. (Note that in this case the i2iDST4 and i2iDST4+RDPCM systems become identical to HEVCv2.) The compression results are presented in Table \ref{tb:res8x8}.

\begin{table}[t]
\setlength{\tabcolsep}{3pt}
\centering
\caption{Average percentage ($\%$) bitrate reduction 
of several systems with minimum allowed block size of 8x8 for intra prediction with respect to HEVCv1 that has minimum allowed block size of 4x4}
\label{tb:res8x8}
\begin{tabular}{l|cccccc}
\hline
          & HEVCv1  & HEVCv2  & i2iDST4\&8   & i2iDST4\&8 \\ 
          &         &         &              & +RDPCM  \\ \hline \hline
          
Class A   & -7.2    & 5.4     &  9.4         & 10.1    \\ \hline
Class B   & -4.4    & 3.3     &  5.3         &  5.7    \\ \hline
Class C   & -6.6    & 2.3     &  3.2         &  4.3    \\ \hline
Class D   & -7.8    & 4.7     &  4.7         &  6.6    \\ \hline
Class E   & -6.7    & 6.4     &  6.3         &  8.3    \\ \hline
Average   & -6.4    & 4.3     &  5.7         &  6.9     \\ \hline
\end{tabular}
\end{table}

Note that the results in Table \ref{tb:res8x8} are bitrate savings with respect to HEVCv1 in the initial set of experiments, i.e. HEVCv1 that has access to all block sizes from 4x4 to 32x32, so that these results can also be easily compared to those in Table \ref{tb:res}. The results in Table \ref{tb:res8x8} indicate that in this new set of experiments, systems employing i2i approximations of ODST-3 achieve similar compression gains with respect to HEVCv2. In particular, the i2iDST4\&8+RDPCM system achieves an average bitrate reduction of $2.7\%$ and $2.6\%$ with respect to HEVCv2 in Tables \ref{tb:res} and \ref{tb:res8x8}, respectively. In summary, from the results in Tables \ref{tb:res} and \ref{tb:res8x8}, it can be concluded that the developed 8-point i2i approximation of the ODST-3 can achieve significant compression gains if it is the smallest point i2i ODST-3 used in the system, but its contribution to the overall compression performance becomes insignificant if it is used together with the 4-point i2i ODST-3.

\subsection{Block size and intra mode statistics}
\label{ssec:stats}
\begin{figure*}[tbh]
\centering
\begin{tikzpicture}
\begin{axis}[
width=\textwidth,
height=5cm, ymin=0, ymax=104,
    symbolic x coords={Cls_A_HEVCv1,Cls_A_HEVCv2,Cls_A_I2I4x4,Cls_A_I2I8x8, s1, 
                       Cls_B_HEVCv1,Cls_B_HEVCv2,Cls_B_I2I4x4,Cls_B_I2I8x8, s2,
                       Cls_C_HEVCv1,Cls_C_HEVCv2,Cls_C_I2I4x4,Cls_C_I2I8x8, s3,
                       Cls_D_HEVCv1,Cls_D_HEVCv2,Cls_D_I2I4x4,Cls_D_I2I8x8, s4,
                       Cls_E_HEVCv1,Cls_E_HEVCv2,Cls_E_I2I4x4,Cls_E_I2I8x8, s5,
                       Cls_V_HEVCv1,Cls_V_HEVCv2,Cls_V_I2I4x4,Cls_V_I2I8x8},
     scaled ticks=false, 
    ybar stacked,
    legend style={at={(0.5,+1.32)}, anchor=north,legend columns=-1},
    scaled x ticks = false,
    x tick label style={rotate=65,anchor=east},
    x label style={at={(axis description cs:0.5,+1.30)},anchor=north},  
      xlabel={ \hspace{-0.0 cm} Class A \hspace{1.05 cm} Class B \hspace{1.05 cm} Class C \hspace{1.05 cm} $~$Class D \hspace{1.05cm} Class E \hspace{1.05cm} Average},
      xtick=data,
      ylabel={Percentage (\%)},
      xticklabels={HEVCv1,HEVCv2,i2iDST4+RDPCM,i2iDST4\&8+RDPCM,  
                   HEVCv1,HEVCv2,i2iDST4+RDPCM,i2iDST4\&8+RDPCM, 
                   HEVCv1,HEVCv2,i2iDST4+RDPCM,i2iDST4\&8+RDPCM, 
                   HEVCv1,HEVCv2,i2iDST4+RDPCM,i2iDST4\&8+RDPCM, 
                   HEVCv1,HEVCv2,i2iDST4+RDPCM,i2iDST4\&8+RDPCM, 
                   HEVCv1,HEVCv2,i2iDST4+RDPCM,i2iDST4\&8+RDPCM}
    ]
\addplot+[ybar] plot coordinates {
(Cls_A_HEVCv1, 93.3) (Cls_A_HEVCv2, 77.3) (Cls_A_I2I4x4, 96.2) (Cls_A_I2I8x8, 78.3) (Cls_B_HEVCv1, 84.0) (Cls_B_HEVCv2, 73.6) (Cls_B_I2I4x4, 83.2) (Cls_B_I2I8x8, 72.6) (Cls_C_HEVCv1, 93.0) (Cls_C_HEVCv2, 83.8) (Cls_C_I2I4x4, 90.7) (Cls_C_I2I8x8, 84.1) (Cls_D_HEVCv1, 98.4) (Cls_D_HEVCv2, 81.1) (Cls_D_I2I4x4, 92.3) (Cls_D_I2I8x8, 86.8) (Cls_E_HEVCv1, 77.9) (Cls_E_HEVCv2, 69.3) (Cls_E_I2I4x4, 76.7) (Cls_E_I2I8x8, 76.3) (Cls_V_HEVCv1, 89.6) (Cls_V_HEVCv2, 77.2) (Cls_V_I2I4x4, 88.1) (Cls_V_I2I8x8, 79.4)  };
\addplot+[ybar] plot coordinates {
(Cls_A_HEVCv1,  6.0) (Cls_A_HEVCv2, 18.2) (Cls_A_I2I4x4,  2.8) (Cls_A_I2I8x8, 20.8) (Cls_B_HEVCv1, 12.5) (Cls_B_HEVCv2, 21.4) (Cls_B_I2I4x4, 12.9) (Cls_B_I2I8x8, 24.6) (Cls_C_HEVCv1,  6.5) (Cls_C_HEVCv2, 13.9) (Cls_C_I2I4x4,  7.7) (Cls_C_I2I8x8, 14.6) (Cls_D_HEVCv1,  1.4) (Cls_D_HEVCv2, 14.3) (Cls_D_I2I4x4,  5.6) (Cls_D_I2I8x8, 11.3) (Cls_E_HEVCv1, 17.5) (Cls_E_HEVCv2, 20.5) (Cls_E_I2I4x4, 15.1) (Cls_E_I2I8x8, 12.7) (Cls_V_HEVCv1,  8.6) (Cls_V_HEVCv2, 17.7) (Cls_V_I2I4x4,  8.7) (Cls_V_I2I8x8, 17.4)  };
\addplot+[ybar] plot coordinates {
(Cls_A_HEVCv1,  0.6) (Cls_A_HEVCv2,  4.6) (Cls_A_I2I4x4,  1.0) (Cls_A_I2I8x8,  0.9) (Cls_B_HEVCv1,  3.5) (Cls_B_HEVCv2,  5.0) (Cls_B_I2I4x4,  4.0) (Cls_B_I2I8x8,  2.8) (Cls_C_HEVCv1,  0.5) (Cls_C_HEVCv2,  2.4) (Cls_C_I2I4x4,  1.6) (Cls_C_I2I8x8,  1.3) (Cls_D_HEVCv1,  0.1) (Cls_D_HEVCv2,  4.6) (Cls_D_I2I4x4,  2.1) (Cls_D_I2I8x8,  1.9) (Cls_E_HEVCv1,  4.6) (Cls_E_HEVCv2, 10.2) (Cls_E_I2I4x4,  8.1) (Cls_E_I2I8x8, 11.0) (Cls_V_HEVCv1,  1.8) (Cls_V_HEVCv2,  5.1) (Cls_V_I2I4x4,  3.2) (Cls_V_I2I8x8,  3.2)   };
\addplot+[ybar] plot coordinates {
(Cls_A_HEVCv1,  0.0) (Cls_A_HEVCv2,  0.0) (Cls_A_I2I4x4,  0.0) (Cls_A_I2I8x8,  0.0) (Cls_B_HEVCv1,  0.0) (Cls_B_HEVCv2,  0.0) (Cls_B_I2I4x4,  0.0) (Cls_B_I2I8x8,  0.0) (Cls_C_HEVCv1,  0.0) (Cls_C_HEVCv2,  0.0) (Cls_C_I2I4x4,  0.0) (Cls_C_I2I8x8,  0.0) (Cls_D_HEVCv1,  0.0) (Cls_D_HEVCv2,  0.0) (Cls_D_I2I4x4,  0.0) (Cls_D_I2I8x8,  0.0) (Cls_E_HEVCv1,  0.0) (Cls_E_HEVCv2,  0.0) (Cls_E_I2I4x4,  0.0) (Cls_E_I2I8x8,  0.0) (Cls_V_HEVCv1,  0.0) (Cls_V_HEVCv2,  0.0) (Cls_V_I2I4x4,  0.0) (Cls_V_I2I8x8,  0.0)  };
\legend{4x4, 8x8, 16x16, 32x32}
\end{axis}
\end{tikzpicture}
\caption{Percentage of pixels that are coded in each available block size in several systems for all sequence classes.}
\label{fig:freq_blksz}
\end{figure*}
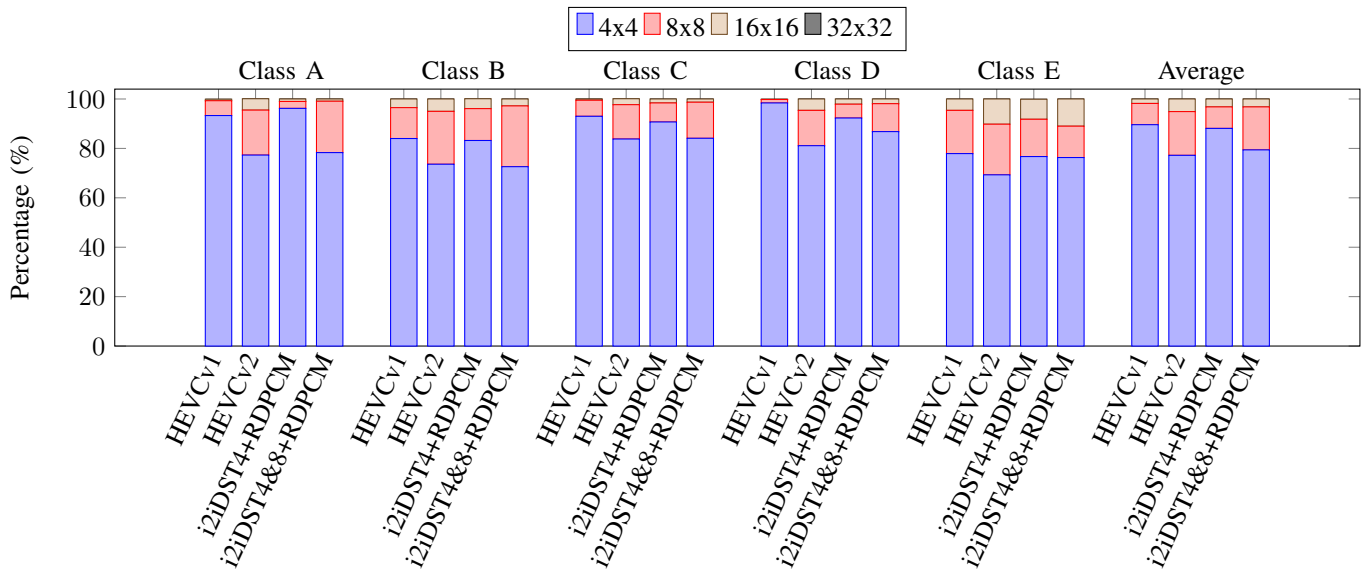

It is also useful to obtain additional insights by looking at the statistics regarding how often available block sizes and intra modes are used in lossless compression within the systems we compare in this section. For this purpose, we use the initial experiments where all block sizes from 4x4 to 32x32 are available for intra prediction in all systems. Figure \ref{fig:freq_blksz} shows the percentage of pixels that are coded in each available block size in the systems HEVCv1, HEVCv2, i2iDST4+RDPCM and i2iDST4\&8+RDPCM for all classes and Table \ref{tb:freq_blksz} summarizes the average of these statistics (averaged over all sequences in all classes.)

The most important observation from the percentages in Figure \ref{fig:freq_blksz} is that 4x4 block size is by far the most frequently used block size in all of the systems for all sequence classes. This is because the bitrate of the prediction residual dominates the overall bitrate in lossless compression (i.e. bitrate of side information, such as intra modes, is a very small fraction of the overall bitrate), and to reduce the bitrate of the residual, better prediction is needed, which is typically best at the smallest available block size, i.e. 4x4 block size. 

A closer look at the percentages in Table \ref{tb:freq_blksz} shows that while the percentages of pixels coded in 4x4 and 8x8 blocks are $89.6\%$ and $8.6\%$ in HEVCv1, respectively, they change to $77.2\%$ and $17.7\%$ in HEVCv2. This is because HEVCv1 does not process the block-based spatial prediction residual and the encoder chooses 4x4 blocks almost exclusively, except in very flat regions where prediction with 4x4 or 8x8 blocks is almost identical. In HEVCv2, the prediction residual is processed with RDPCM in horizontal and vertical intra modes, which improves the prediction performance in larger (and smaller) blocks and thus larger blocks are used more often in HEVCv2. 

\begin{table}[b]
\setlength{\tabcolsep}{3pt}
\centering
\caption{Percentage of pixels that are coded in each available block size in several systems (average over all sequences)}
\label{tb:freq_blksz}
\begin{tabular}{l|ccccc}
\hline
Block size & HEVCv1 & HEVCv2  & i2iDST4  & i2iDST4\&8  \\ 
        &        &         & +RDPCM   & +RDPCM      \\ \hline \hline
4x4     & 89.6   & 77.2    & 88.1     & 79.4 \\ \hline
8x8     & 8.6    & 17.7    & 8.7      & 17.4 \\ \hline
16x16   & 1.8    & 5.1     & 3.2      & 3.2 \\ \hline
32x32   &   0    & 0       & 0        & 0 \\ \hline
\end{tabular}
\end{table}

\begin{table}[b]
\setlength{\tabcolsep}{3pt}
\centering
\caption{Distribution of percentages of the 4x4 and 8x8 blocks in Table \ref{tb:freq_blksz} to horizontal\&vertical and remaining intra modes} 
\label{tb:freq_modes}
\begin{tabular}{l|ccccc}
\hline
Intra modes  & HEVCv1 & HEVCv2  & i2iDST4 & i2iDST4\&8  \\ 
             &        &         & +RDPCM  & +RDPCM      \\ \hline \hline
4x4 hor\&ver & 15.0   & 41.2    & 25.9    & 23.9 \\ \hline
4x4 other    & 74.6   & 36.0    & 62.2    & 55.5 \\ \hline
8x8 hor\&ver & 0.6    & 12.0    & 4.5     & 5.1 \\ \hline
8x8 other    & 8.0    & 5.7     & 4.2     & 12.3 \\ \hline
\end{tabular}
\end{table}

Let us also observe what these percentages are in the systems utilizing i2i approximations of the ODST-3. In the i2iDST4+RDPCM system, the percentage of pixels coded in 4x4 blocks increases back to $88.1\%$ while the percentage of pixels coded in 8x8 blocks decreases to $8.7\%$. This change is due to the i2i ODST-3 in 4x4 blocks, which improves the lossless compression performance in 4x4 blocks and thus the encoder chooses 4x4 blocks more often. In the i2iDST4\&8+RDPCM system, the percentage of pixels coded in 4x4 blocks decreases back to $79.4\%$, while the percentage of pixels coded in 8x8 blocks increases back to $17.4\%$. This change is due to using i2i ODST-3 also in 8x8 blocks, which can slightly improve lossless compression performance in 8x8 blocks and thus the encoder chooses 8x8 blocks more often.

Table \ref{tb:freq_modes} shows how the percentages of the 4x4 and 8x8 blocks in Table \ref{tb:freq_blksz} are distributed to the 35 intra modes. In particular, we consider the aggregate of the horizontal and vertical intra modes 
and the the aggregate of the remaining intra modes. Table \ref{tb:freq_modes} shows that in HEVCv1, $15.0\%$ of pixels are coded in 4x4 block horizontal and vertical intra modes and $74.6\%$ in the remaining modes, and these numbers change to $41.2\%$ and $36.0\%$ in HEVCv2. This change  happens because HEVCv2 uses RDPCM in horizontal and vertical modes and RDPCM improves compression performance, causing the encoder to choose the modes more often. In the i2iDST4+RDPCM system, i2i ODST-3 is used in the 4x4 block other modes (i.e. not horizontal or vertical) and this increases the percentage of these modes to $62.2\%$ since the i2i ODST-3 improves compression performance and thus the encoder chooses these modes more often. In the i2iDST4\&8+RDPCM system, i2i approximation of ODST-3 is also used in the other modes of 8x8 blocks and this increases the percentage of the 8x8 other modes to $12.3\%$ compared to the $4.2\%$ in the i2iDST4+RDPCM or the $5.7\%$ in the HEVCv2 systems, which do not process these intra modes prior to entropy coding.

\section{Conclusions}
\label{sec:conc}
This paper explored an alternative approach for lossless intra-frame compression. A popular and computationally efficient approach, used also in H.264 and HEVC, is to skip transform and quantization but also process the residual block with DPCM, along he horizontal or vertical direction, prior to entropy coding. This paper explored an alternative approach based on processing the residual block with integer-to-integer (i2i) transforms. In particular, we developed novel i2i approximations of the odd type-3 DST (ODST-3) that can be applied to the residuals of all intra prediction modes in lossless intra-frame compression. Experimental results with the HEVC reference software showed that the developed i2i approximations of the ODST-3 improve lossless intra-frame compression efficiency with respect to HEVC version 2, which uses the popular DPCM method along the horizontal or vertical direction, by an average 2.7\% without a significant effect on computational complexity.



\section*{Acknowledgment}
We thank Vivek K Goyal for his valuable comments.

\ifCLASSOPTIONcaptionsoff
  \newpage
\fi



\bibliographystyle{IEEEtran}
\bibliography{IEEEabrv,refs}

%

%

\begin{IEEEbiography}[{\includegraphics[width=1in,height=1.25in,clip,keepaspectratio]{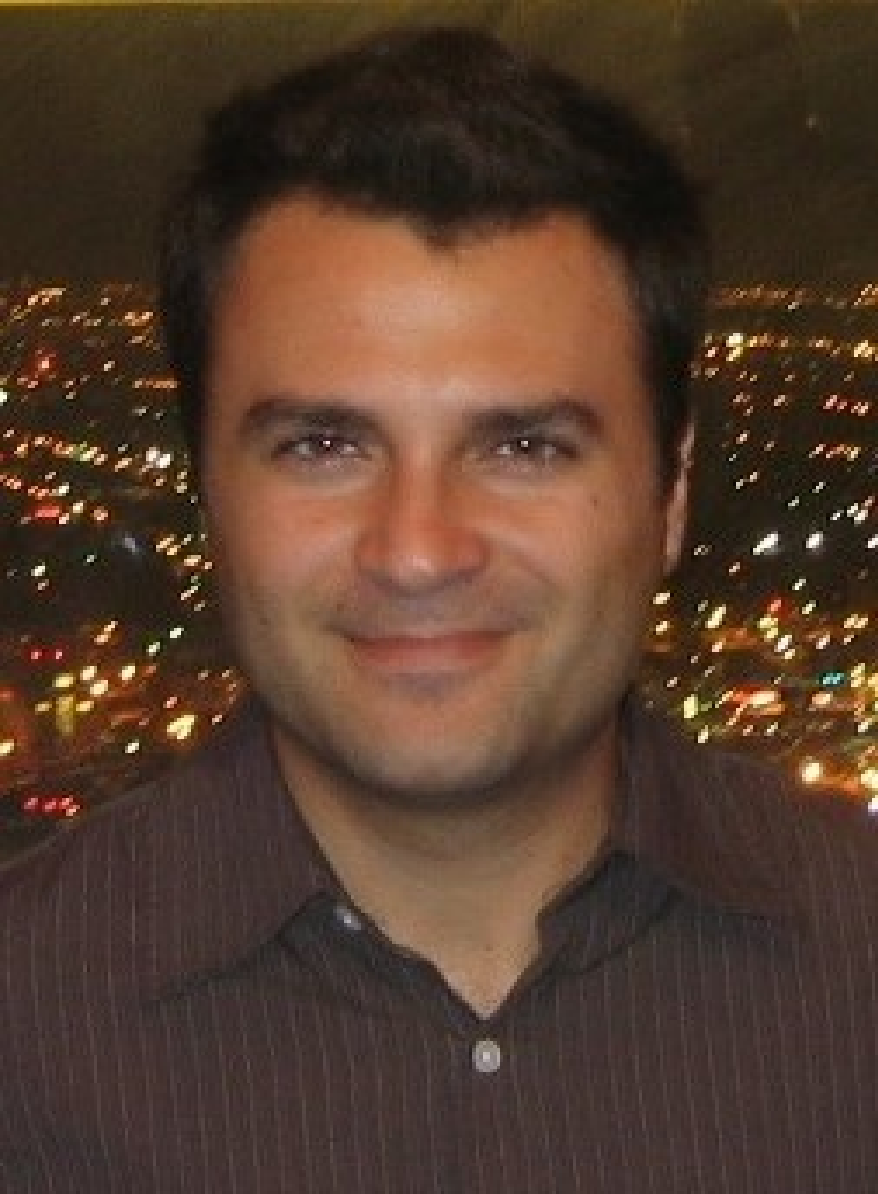}}]{Fatih Kamisli}
(S'09-M'11) received the B.S. degree from the Middle East Technical University, Ankara, Turkey, in 2003, and the M.S. and Ph.D. degrees in Electrical Engineering and Computer Science from the Massachusetts Institute of Technology, Cambridge, MA, USA in 2006 and 2010, respectively.

He is an Assistant Professor in the Electrical and Electronics Engineering Department at the Middle East Technical University, Ankara, Turkey. His current research interests include image and video processing, in particular, compression.
\end{IEEEbiography}







\end{document}